\documentclass[superscriptaddress,reprint,aps,pre,nolongbibliography]{revtex4-2}
\usepackage{amsmath,amsfonts}
\usepackage{graphicx}
\usepackage{amssymb}
\usepackage{enumerate}
\usepackage{mathrsfs}
\usepackage{mathtools}
\usepackage{float}
\usepackage[colorlinks=true,linkcolor=blue,citecolor=blue,linkcolor=blue]{hyperref}

\begin{document}
\title{Further results on the
power-law decay of the fraction of the mixed eigenstates in kicked top model with mixed-type classical phase space}
\author{Hua Yan}
\email{yanhua@ustc.edu.cn}
\affiliation{CAMTP - Center for Applied Mathematics and Theoretical Physics,    University of Maribor, Mladinska 3, SI-2000 Maribor, Slovenia}
\author{Qian Wang}
\affiliation{CAMTP - Center for Applied Mathematics and Theoretical Physics,    University of Maribor, Mladinska 3, SI-2000 Maribor, Slovenia}
\affiliation{Department of Physics, Zhejiang Normal University, Jinhua 321004, China}
\author{Marko Robnik}
\email{Robnik@uni-mb.si}
\affiliation{CAMTP - Center for Applied Mathematics and Theoretical Physics,    University of Maribor, Mladinska 3, SI-2000 Maribor, Slovenia}
\date{\today} 
\begin{abstract}
By using the Krylov subspace technique to generate the spin coherent states in kicked top model, a prototype model for studying quantum chaos, the accessible system size for studying the Husimi functions of eigenstates can be much larger than that reported in the literature and our previous study \href{https://journals.aps.org/pre/abstract/10.1103/PhysRevE.108.054217}{Phys. Rev. E 108, 054217 (2023)}. In the fully chaotic kicked top, we show that the mean Wehrl entropy localization measure approaches the prediction given by the Circular Unitary Ensemble, while the spectral statistics follows the Circular Orthogonal Ensemble. For the mixed-type cases, we identify mixed eigenstates by the overlap of the Husimi function with regular and chaotic regions in classical compact phase space. Numerically, we show that the fraction of mixed eigenstates scales as $j^{-\zeta}$, a power-law decay as the system size $j$ increases, across nearly two orders of magnitude, expanding significantly beyond the previously limited range of system sizes. This provides supporting evidence for the principle of uniform semiclassical condensation of Husimi functions and the Berry-Robnik picture in the semiclassical limit.
\end{abstract}  
\maketitle
\section{Introduction}
At the very beginning of quantum chaos it was Percival \cite{percival1973regular} who proposed to distinguish between the regular quantum eigenstates and the chaotic ones, associated with classical regular and chaotic regions in the phase space, respectively. They should be distinguished by the nature of their energy spectra, namely by their sensitivity with respect to small perturbations. This line of thought led to the semiclassical eigenfunction hypothesis by Berry \cite{berry1977regular}, which is crucial for understanding quantum eigenstates. It proposes that in the semiclassical limit, quantum eigenstates concentrate on compact phase space regions explored by typical trajectories in the long time limit. In integrable systems, these are the invariant tori, while in ergodic dynamics, eigenstates equidistribute on the energy surface. In mixed-type systems with both regular and chaotic motions in phase space \cite{lichtenberg2013regular}, eigenstates concentrate either on regular or chaotic regions.  This hypothesis finally develops into the so-called principle of uniform semiclassical condensation (PUSC) of Wigner functions (or Husimi functions) \cite{robnik1998topics}. This principle, in turn, directly leads to the eigenstate thermalization hypothesis conjecture by Srednicki \cite{srednicki1994chaos}, which is now recognized as an explanation for the emergence of thermodynamic equilibrium in isolated quantum many-body systems \cite{rigol2008thermalization,d2016quantum,wang2022eigenstate}. Moreover, as a consequence of this principle, the spectral statistics for regular and chaotic states in the semiclassical limit are separately described by Poissonian statistics \cite{berry1977level} and random matrix theory \cite{casati1980connection,bohigas1984characterization,muller2004semiclassical}, while the entire spectrum is effectively captured by the Berry-Robnik picture \cite{berry1984semiclassical}.

Regarding the behavior of quantum eigenstates in mixed-type systems away from the semiclassical limit, one expects various tunneling processes between different phase-space structures. Among these are the  chaos-assisted tunneling between different regular regions \cite{tomsovic1994chaos,bohigas1993manifestations,doron1995semiclassical}, the flooding of chaotic states into regular islands \cite{backer2005flooding}, and the so-called resonance-assisted tunneling between the regular island across the chaotic sea \cite{brodier2001resonance,backer2008regular,backer2008dynamical,lock2010regular}. The general behavior of quantum states  approaching the semiclassical limit is not well studied \cite{ketzmerick2000new}.  Through the overlap of the Husimi function with both regular and chaotic regions in classical compact phase space, mixed eigenstates can be identified, distinct from regular and fully chaotic states. Very recently, in quantum billiards \cite{lozej2022phenomenology}, it was found that the fraction of mixed states shows a power-law decay as the semiclassical limit is approached, across two orders of magnitude. In quantum kicked top (QKT), a prototype Floquet model for studying quantum chaos \cite{haake2013quantum,haake1987classical}, our previous study \cite{wang2023power} also showed a power-law decay for mixed-type QKT, but only within a relatively small range of system sizes. This limitation arises from our method used to generate the spin coherent state (SCS), where a direct calculation of large factorials is unavoidable. 

In this paper, we use the Krylov subspace techniques \cite{sidje1998expokit,moler2003nineteen,golub2013matrix,saad2003iterative} to generate the SCS and avoid the large factorials. This allows for a significantly larger accessible system size for studying Husimi functions, compared to that reported in literature and our previous study. It is important to note that the issue of large factorials in SCS is similar to the longstanding problem encountered in the numerical evaluation of the Wigner rotation matrix, which is crucial in modern nuclear-structure models. In both cases, directly calculating large factorials at high spins results in significant precision loss, instability, and increased memory consumption \cite{tajima2015analytical,wang2022effective}, as SCS can be derived from a SU(2) rotation. In the fully chaotic QKT, by using SCSs generated through the Krylov subspace method, we find that the mean Wehrl entropy localization measure (ELM) approaches the prediction given by the Circular Unitary Ensemble (CUE), provided that coherent states situated on symmetrical lines exhibit COE eigenvector statistics but have zero measure in phase space, whereas those outside follow CUE eigenvector statistics. For the mixed-type cases, we show a power-law decrease of the fraction of mixed eigenstates as system size increases over nearly two orders of magnitude, corroborating the PUSC and the Berry-Robnik picture in the semiclassical limit.

The paper is organized as follows. In Sec. \ref{sec2}, we
introduce QKT and analyze the transition to chaos for both its classical correspondence and quantum case. In Sec. \ref{sec3}, we give the definition of the SCS and apply the Krylov subspace technique to generate them. We then establish a consistent transition to chaos using the mean Wehrl ELM, compared to other indicators of chaos. In the fully chaotic QKT, we study the mean Wehrl ELM versus the system size. In Sec. \ref{sec4}, we analyze the joint distribution of the phase-space overlap index and Wehrl ELM, and show as the system size increases, how the distribution of the former evolves, as well as the fraction of mixed states. We draw our conclusion in Sec. \ref{sec5}.

\section{Quantum kicked top}
\label{sec2}
The QKT is described by the Hamiltonian \cite{haake2013quantum}
 \begin{align}
    \label{eq:KT-Hamiltonian}
    H=\alpha J_x +\frac{\gamma}{2j}J_z^2\sum_{n=-\infty}^{\infty}\delta(t-n),
\end{align}
where the dynamical variables of the top are three components of the angular momentum operators of the spin-$j$ system, can also be expressed in terms of $2j$ collective spin-$\frac{1}{2}$ Pauli operators, for example $J_z=\sum_{k=1}^{2j}\sigma_z^{(k)}/2$. The dimension of the Hilbert space is $N=2j+1$, and the squared angular momentum is conserved, $J^2=j(j+1)$ with $j$ integer or half-integer (in this work, we consider integer $j$). The first term in Eq. \eqref{eq:KT-Hamiltonian} describes a precessional rotation about the $x$-axis with angular frequency $\alpha$, the second term denotes a torsional rotation around the $z$-axis with strength $\gamma$ (with the period set to unity). 

The dynamical evolution of the QKT is governed by the Floquet operator
\begin{align}
    F = \exp(-i\frac{\gamma}{2j}J_z^2)\exp(-i\alpha J_x),
\end{align}
is invariant under rotation by the angle $\pi$ along the $x$-axis $R_x=e^{i\pi J_x}$, which suggests the even or odd parity of the quasi-energy. Using this symmetry, the $(2j+1)$-dimensional unitary matrix $F$ in the representation of the Dicke states \{$|j,m\rangle, |m|\le j$\}, i.e. eigenstates of $J_z$, can further be reduced to two unitary matrices 
\begin{align}
\label{eq:floquet-parity}
  F_{m,m'}^{\pm }&=\langle j,m,\pm|F|j,m',\pm\rangle,
\end{align}
where  
\begin{align}
    \label{eq:parity-basis}
    |j,m,\pm\rangle=(|j,m\rangle\pm|j,-m\rangle)/\sqrt{2},
\end{align}
with $1\le m \le j$ for the odd parity, and $0\le m\le j$ for the even parity where $|j,0,+\rangle$ is not normalized.
It should be noted that at $\alpha=\pi/2$, the QKT has a further symmetry \cite{haake1987classical}. In this paper, we set $\alpha=11\pi/19$, the same as the previous work.  
 
The Heisenberg equation of motion is given by the map $J_{i,n+1}=F^\dagger J_{i,n}F$, where $J_{i,n}\equiv J_i(n)$ denotes the time-evolved operators at $t=n$. Using Baker-Campbell-Hausdorff (BCH) formula for the expansion of the map, one has 
\begin{align}
    \label{eq:Heisenberg}
    J_{x,n+1}&=\frac{1}{2}[\tilde{J}_{x,n}+i\tilde{J}_{y,n}]\exp\left[i\frac{\gamma}{2j}(2\tilde{J}_{z,n}+1)\right]+h.c., \nonumber \\
   J_{y,n+1}&=\frac{1}{2i}[\tilde{J}_{x,n}+i\tilde{J}_{y,n}]\exp\left[i\frac{\gamma}{2j}(2\tilde{J}_{z,n}+1)\right]+h.c, \nonumber \\
   J_{z,n+1}&=\tilde{J}_{z,n},
\end{align}
where $\tilde{J}_{x,n} = J_{x,n}$,  $\tilde{J}_{y,n}=J_{y,n}\cos \alpha-J_{z,n}\sin \alpha$, and $\tilde{J}_{z,n} = J_{y,n}\sin \alpha + J_{z,n}\cos \alpha$. 

\subsection{The transition to Hamiltonian chaos}
The classical map emerges in the classical limit as $j\to \infty$ where noncummutivity is negligible and the rescaled Heisenberg operators $\mathbf{x}_n\equiv \mathbf{x}(n)=\mathbf{J}_n/j$ tend to a unit vector with commuting components. From Eq. \eqref{eq:Heisenberg}, the classical equations of motion follow
\begin{align}
    \label{eq:classical-map}
  x_{n+1}&=\tilde{x}_n\cos \gamma \tilde{z}_{n}-\tilde{y}_n\sin\gamma \tilde{z}_{n}, \nonumber \\ 
   y_{n+1}&=\tilde{x}_n\sin\gamma \tilde{z}_{n}+\tilde{y}_n\cos\gamma \tilde{z}_{n}, \nonumber\\
    z_{n+1}&=\tilde{z}_{n},
\end{align}
where $\tilde{x}_n=x_n$,  $\tilde{y}_n=y_n\cos \alpha-z_n\sin \alpha$, and $\tilde{z}_n=y_n\sin \alpha+z_n\cos \alpha$. 
This classical map $\mathbf{x}_{n+1}=F_c(\mathbf{x}_{n})$ is restricted on the unit sphere $\mathbb{S}^2$ and the Jacobian matrix $\partial F_c/\partial \mathbf{x}_n=\Omega_n\cdot R_{\alpha}$, where
\begin{align}
    \Omega_n&=\begin{pmatrix}
      \cos\gamma \tilde{z}_{n},& -\sin\gamma \tilde{z}_{n},&  -\gamma\tilde{x}_n\sin \gamma \tilde{z}_{n}-\gamma\tilde{y}_n\cos\gamma \tilde{z}_{n} \\
      \sin\gamma \tilde{z}_{n},& \cos\gamma \tilde{z}_{n},  &   \gamma\tilde{x}_n\cos\gamma \tilde{z}_{n}-\gamma\tilde{y}_n\sin\gamma \tilde{z}_{n}\\
      0& 0& 1
      \end{pmatrix}, \nonumber \\
    R_\alpha&=\begin{pmatrix}
        1& 0&  0  \\
      0& \cos \alpha &  -\sin \alpha\\
       0& \sin \alpha & \cos \alpha
        \end{pmatrix}.
  \end{align}

  \begin{figure}
    \includegraphics[width=1.0\linewidth]{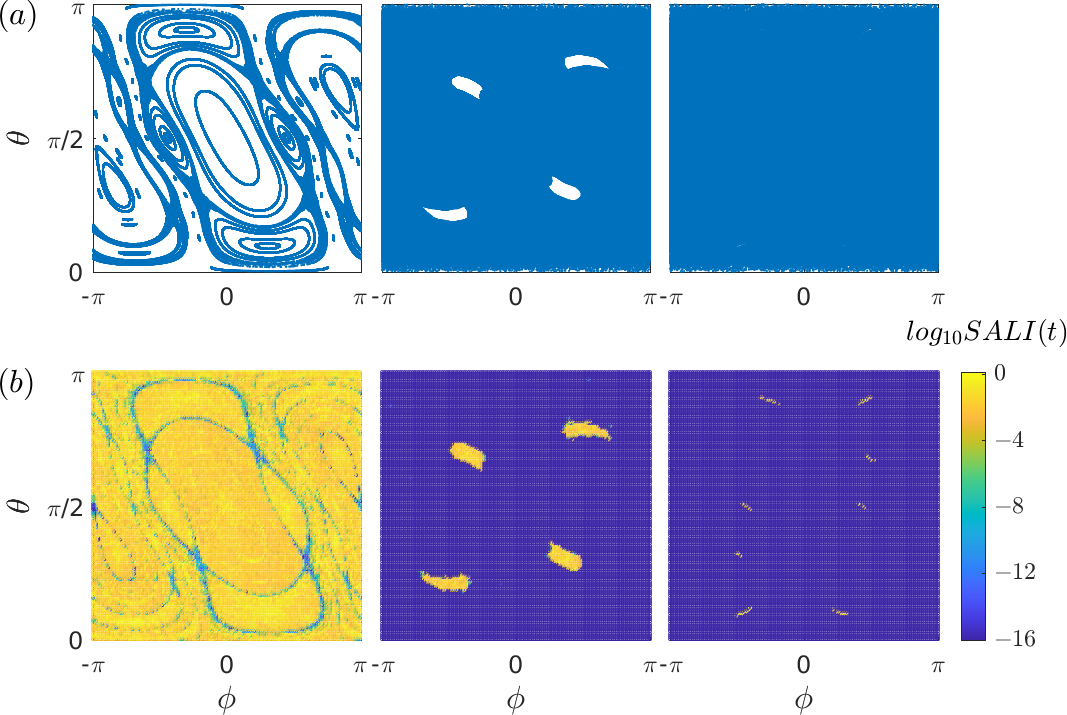}
    \caption{(a) A stroboscopic map of the classical dynamics generated from 20 random initial conditions of $5\times10^4$ iterations (top panels). (b) The corresponding logarithmic values of SALI (bottom panels) on the parametrized phase space $(\theta,\phi)$ discretized by $200\times 400$ grids of square cells, of same area, after 300 kicks. The
    initial conditions colored dark blue correspond to chaotic orbits, the yellowish indicates ordered motion, and the intermediate suggests sticky orbits. From left to right, the kicking strength $\gamma=2, 4, 6$ and $
    \alpha=11\pi/19$.}
    \label{fig:saliKT}
\end{figure}

 The stroboscopic dynamics of the classical kicked top (CKT) is symplectic since $\det(\partial F_c/\partial \mathbf{x}_n)=1$, and $\mathbb{S}^2$ as the phase space of the classical top can be parametrized as: $x=\sin\theta\cos\phi$, $y=\sin\theta\sin\phi$, $z=\cos\theta$, with $z$ and $\phi$ being the canonical pair. Two symmetries of CKT can be found by two involutions \cite{haake1987classical}, defined by
 \begin{subequations}
    \label{eq:time-ckt}
 \begin{alignat}{2}
    T_1^c(\mathbf{x})&=(x,-y\cos\alpha +z\sin\alpha,y \sin\alpha+z\cos\alpha),\\
    T_2^c(\mathbf{x})&=(x, y\cos\alpha -z\sin\alpha,-y\sin\alpha-z\cos\alpha),
 \end{alignat}
\end{subequations}
where $(T_1^c)^2=(T_2^c)^2=1$ with the determinants $\det(\partial T_1^c/\partial \mathbf{x})=\det (\partial T_2^c/\partial \mathbf{x})=-1$. These two involutions yield time reversal operations for the classical map $F_c$ in the sense that $T_1^cF_cT_1^c=T_2^cF_cT_2^c=F_c^{-1}$, indicating that the map has double reversible dynamics.

 CKT undergoes a transition from integrability to chaos with increasing kicking strength $\gamma$, as illustrated in Fig. \ref{fig:saliKT}(a), the Poincar\'e section of the classical map, for different values of $\gamma$. In the classical phase space, regular orbits predominate at small values of $\gamma$. As the kicking strength increases, mixed dynamics emerge, characterized by the coexistence of regular orbits and chaotic motion. Eventually, for larger $\gamma$, the system transitions into full chaos, as illustrated in the rightmost panel of Fig. \ref{fig:saliKT}(a).

 To qualitatively demonstrate the transition to chaos in the classical map, in this work, we use the smaller alignment index (SALI) \cite{skokos2001alignment,skokos2003does,skokos2004detecting} which rely on the evolution of deviation vectors from a given orbit, to detect regular and chaotic motion of the symplectic map. For chaotic orbits, it was proven that SALI$(t)\propto e^{(-L_1-L_2)t}$ \cite{bountis2012complex}, with $L_1,L_2$ being the two largest Lyapunov exponents (LEs). While in three-dimensional symplectic map $L_2=0$, therefore for CKT, the SALI is directly related to the largest LE as SALI$(t)\propto e^{-L_1t}$.  Fig. \ref{fig:saliKT}(b) illustrates the logarithmic SALI values after 300 kicks for grid points, at each point plotted using an assigned color according to its value. Upon comparing with the corresponding Poincar\'e sections in Fig. \ref{fig:saliKT}(a), in addition to the resemblance, we also acquire a much more detailed
 picture of the regions where chaotic or regular motion
 occurs, and at the borders between these regions we find
 intermediate colors which correspond to sticky orbits. 
 
\begin{figure}
    \includegraphics[width=0.9\linewidth]{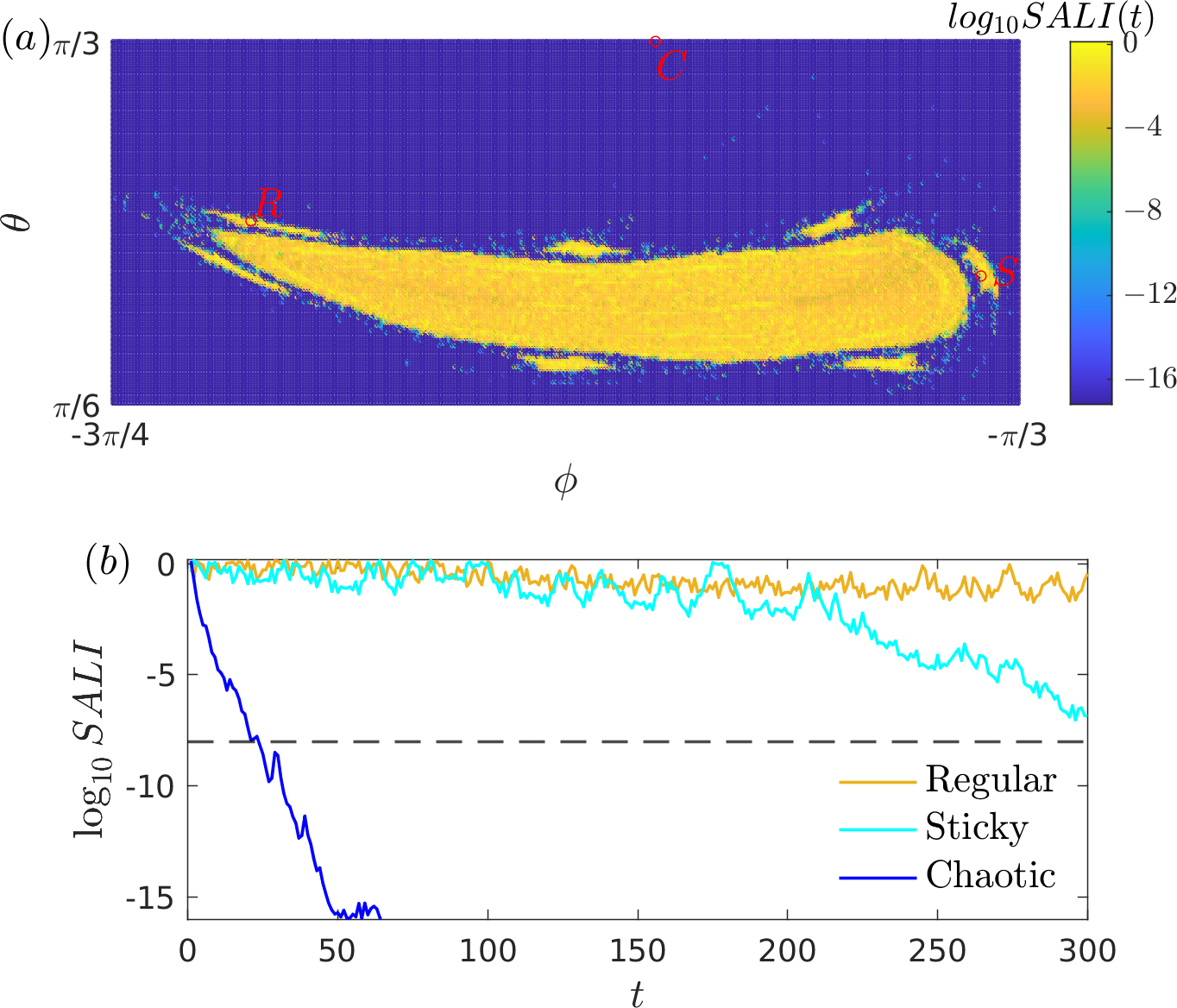}
    \caption{(a) Magnified view of the SALI (after 300 kicks) plot for the CKT with $\gamma=4$, highlighting the left-bottom region around the regular island in the middle panel of Fig. \ref{fig:saliKT}(b). (b) Evolution of the SALI with respect to time t, for three
    initial conditions chosen from panel (a), denoted by
    R (regular), S (sticky) and C (chaotic). The SALI of sticky orbit takes a longer time to approach the threshold value $10^{-8}$, as it remains near the border between the chaotic and regular regions up to approximately $t=200$.}
    \label{fig:sali-details}
\end{figure}

Fig. \ref{fig:sali-details}(a) shows a magnified SALI plot of the phase space, for the CKT at $\gamma=4$, focusing on the region around the regular island depicted in the middle panel of Fig. \ref{fig:saliKT} (b). Clearly, smaller islands of regular motion appear around the main island, organized in a hierarchical way \cite{lichtenberg2013regular}. In addition, along the border between the regular islands and the chaotic sea, the existence of stickiness is evident, indicated by intermediate SALI values.  In deterministic dynamical systems, stickiness is associated with the partial barriers, such as remnants of destroyed invariant tori \cite{meiss2015thirty}, which slow down the transport between different regions of phase space. Fig. \ref{fig:sali-details}(b) illustrates the typical time evolution of the SALI for three types of orbits: regular, sticky, and chaotic. It shows that the sticky orbit remains near the border between the chaotic and regular region for a very long time, requiring significantly more iterations than typical chaotic orbits to reveal its true chaotic behavior.

To further quantify the degree of chaos, we define the chaotic fraction $\mu_c$ as the relative area of the chaotic part of the phase-space as 
\begin{align}
    \label{eq:mu-c}
   \mu_c=\frac{1}{4\pi}\int \chi_c(\theta,\phi) \sin\theta d\theta d\phi,
\end{align}
where $\chi_c(\theta,\phi)$ denotes the characteristic function of the chaotic component, which takes the value of 1 on chaotic region and zero otherwise. The chaotic fraction $\mu_c$ can be used as an indicator of chaos. It measures the transition from integrable dynamics with $\mu_c$ = 0 to the fully chaotic (ergodic) $\mu_c=1$. Our criterion for the
classification of initial conditions belonging to a chaotic
region is that SALI$(t)\le 10^{-8}$ at $t = 300$, implying that the deviation vectors have been aligned. Calculated from $200\times 400$ initial grid points on the parametrized phase space $(\theta,\phi)$, Fig. \ref{fig:FractionRatio}(a) illustrates the dependence of the chaotic fraction $\mu_c$ on $\gamma$, for CKT. The value of $\mu_c$ remains near 0 for $\gamma \leq 2$ and increases thereafter until it saturates to 1 for $\gamma \gtrsim 4.4$. It should be noted that the transition point at $\gamma \simeq 2$ obtained by the chaotic fraction $\mu_c$ from the SALI method is consistent with the transition point from the Kolmogorov-Sinai entropy given in Ref. \cite{wang2023power}.

\subsection{The transition to quantum chaos}
\label{sec2.2}
\begin{figure}
    \includegraphics[width=0.9\linewidth]{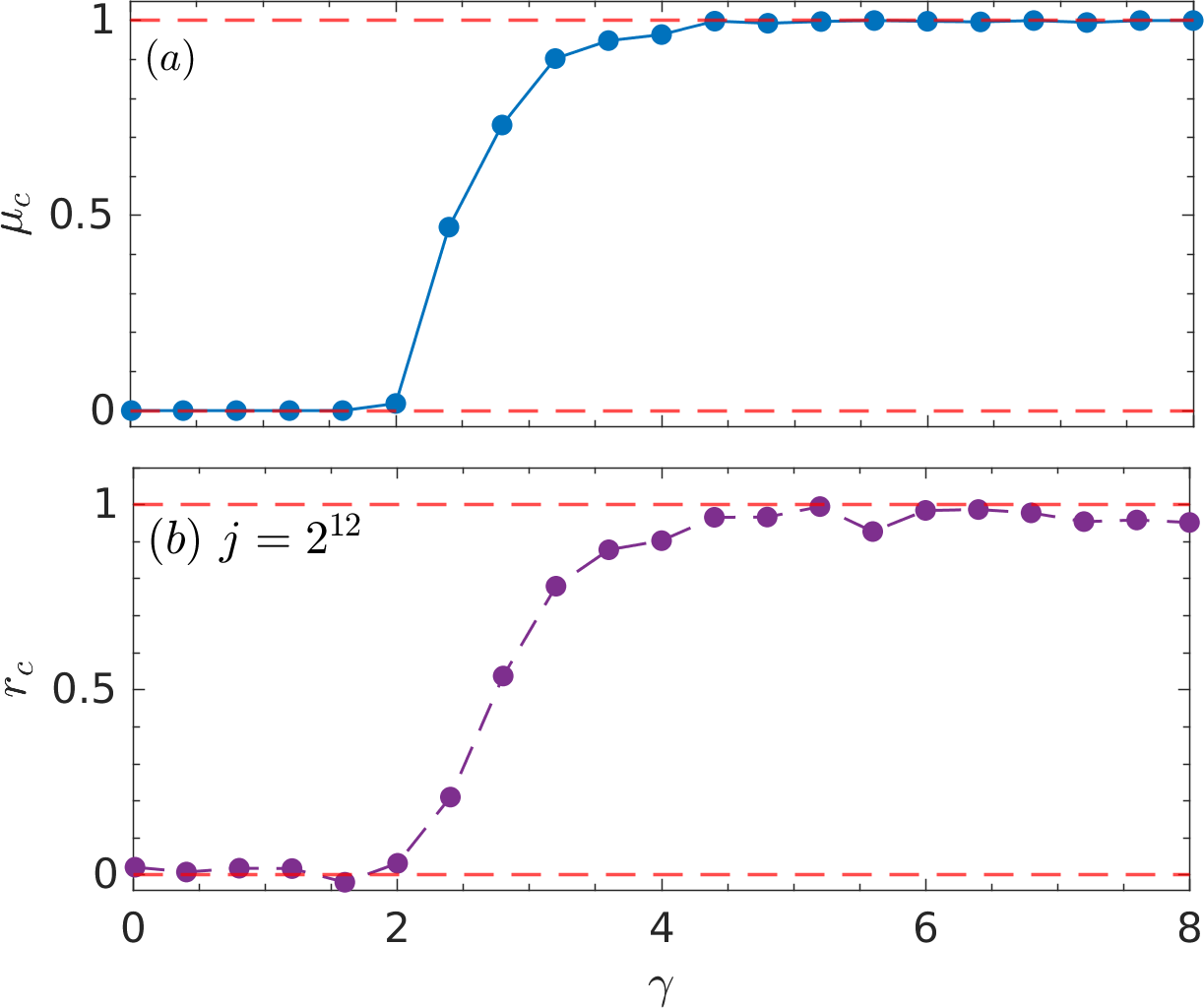}
    \caption{(a) Chaotic fraction $\mu_c$ of the classical phase space of CKT,  as a function of $\gamma$, calculated from $200\times 400$ initial grid points on the parametrized phase space $(\theta,\phi)$. (b) The normalized mean spacing ratio $r_c$ averaged over the odd and even parity, versus $\gamma$ for QKT with $j=2^{12}$.  In both figures, the lower and upper (red) dashed lines denote the value $0$ and $1$, and the parameter $\alpha=11\pi/19$.}
    \label{fig:FractionRatio}
\end{figure}

The transition to Hamiltonian chaos, as indicated by the variation of $\mu_c$ versus the kicking strength in Fig. \ref{fig:FractionRatio}(a),  would also be manifested in the quantum realm due to the quantum-classical correspondence. There are several methods to diagnose quantum chaos, statistically or dynamically. One is the short-range statistical properties of spectra, particularly the nearest-neighbor-level spacing distribution, denoted as $P(s)$, where $s$ denotes the spacing between two unfolded consecutive energy levels.  In the quantum integrable case, this distribution follows the Poisson distribution $P(s)=e^{-s}$, whereas the quantum chaotic case with the time-reversal symmetry is the well-known Wigner-Dyson distribution $P(s)=(\pi/2)s\exp(-\pi s^2/4)$ from the random-matrix statistics. 

To avoid the numerical spectral unfolding, we consider instead the spacing ratios, defined as \cite{oganesyan2007localization,atas2013distribution}
\begin{align}
  r_n = \min(\frac{s_n}{s_{n-1}},\frac{s_{n-1}}{s_n}),
\end{align}
with $s_n=v_n-v_{n-1}$ the consecutive level spacing, from $\{v_n\}$ an ordered set of quasienergy levels, $F|v_n\rangle= e^{iv_n}|v_n\rangle$. The approximate formulas for the distribution of \( P(r) \) can be derived from random matrix statistics. For the Poisson level spacing, the resulting mean spacing ratio $ \langle r \rangle$ is $ \langle r \rangle_p = 2\ln2 - 1\approx 0.386$, whereas for Circular Orthogonal Ensemble (COE) statistics, $ \langle r \rangle_{COE}\approx 0.536$ \cite{atas2013distribution,d2014long}. We can then define a quantity $r_c$ as the normalized mean spacing ratio, as an indicator of quantum chaos, namely
\begin{align}
r_c = \frac{\langle r \rangle-\langle r \rangle_P}{\langle r \rangle_{COE}-\langle r \rangle_P}.
\end{align}
Clearly, $r_c=0$ corresponds to the integrable case, while $r_c=1$ signifies full chaos. As discussed in Ref. \cite{haake1987classical}, in QKT, there exist two antiunitary time reversal operators
\begin{align}
    \label{eq:t1t2}
    T_1=e^{i\alpha J_x}e^{i\pi J_z}K,\quad T_2=e^{-i\alpha J_x}e^{i\pi J_y}K,
\end{align}
 where $K$ denotes the conjugation in the eigenbasis of $J_z$. These operators satisfy $T_1^2=T_2^2=1$ and time reversal property $T_1FT_1=T_2FT_2=F^\dagger$. One can therefore
expect that in the fully chaotic regime, the statistics of quasienergy of $F$ must be given by COE. As depicted in Fig. \ref{fig:FractionRatio}(b), the variation of $r_c$, averaged from both parities, with kicking strength $\gamma$ illustrates that it remains close to 0 for $\gamma \leq 2$, gradually increasing thereafter until it reaches a plateau near 1 for $\gamma \geq 4.4$. This behavior precisely aligns with the chaotic fraction $\mu_c$ versus $\gamma$ depicted in Fig. \ref{fig:FractionRatio}(a), as well as with the mean Wehrl entropy localization length of eigenstates, which we will demonstrate in the following.

\section{Husimi function and Localization of eigenstates}
\label{sec3}
By employing spectral statistics, we have demonstrated in Sec. \ref{sec2.2} the transition to quantum chaos, which aligns with the classical route to chaos. In this section, we introduce the SCS and employ the Krylov subspace technique to numerically calculate the SCS. Notably, by avoiding the problem of large factorials, the accessible value of $j$ can be several orders of magnitude higher than those reported in the literature, including our previous work \cite{wang2023power}. Following this, by analyzing the Husimi function of eigenstates based on SCS, we can explore the localization of these eigenstates, especially the Wehrl entropy \cite{wehrl1978general} and the corresponding entropy localization measure (ELM), to a deeper semiclassical regime.

\subsection{Spin coherent state and Husimi functions}
The spin coherent state $|\theta,\phi\rangle$ from SU(2) \cite{perelomov1972coherent,klauder1985coherent,zhang1990coherent}, can be obtained as a rotation on the Dicke manifold, expanded in the Dicke basis $|j, m\rangle$ as
\begin{subequations}
    \begin{align}
        |\theta&,\phi\rangle =R(\theta,\phi)|j,j\rangle= \exp(\mu J_--\mu^*J_+)|j,j\rangle \label{scs-subeq1}\\ 
        &= (1+|\xi|^2)^{-j}\sum_{m=-j}^{j}\binom{2j}{j+m}^{1/2}\xi^{j-m}|j,m\rangle \label{scs-subeq2},
    \end{align} 
\end{subequations}
where $\mu=\frac{\theta}{2}\exp(i\phi)$, $\xi=\tan(\frac{\theta}{2})\exp(i\phi)$ with $\theta\in[0,\pi]$ and $\phi\in[-\pi,\pi)$, $J_\pm=J_x\pm iJ_y$.  The SU(2) generalized coherent states $|\theta,\phi\rangle$ are overcomplete, so that
\begin{align}
    \frac{2j+1}{4 \pi}\int \sin\theta d\theta d\phi\ |\theta,\phi\rangle \langle\theta,\phi| = \mathbb{I}.
\end{align} 
The Husimi function of the eigenstate $|v_n\rangle$ of the Floquet operator $F$ is then given by
\begin{align}
    Q_n(\theta,\phi)=|\langle \theta,\phi|v_n\rangle|^2,
\end{align}
with the normalization condition
\begin{align}
    \label{eq:norm2}
    \frac{2j+1}{4\pi}\int \sin\theta d\theta d\phi\ Q_n(\theta,\phi)=1.
\end{align}

Numerical studies of Husimi functions in the past usually employed Eq. \eqref{scs-subeq2} for the SCS representation of quantum states. However, the presence of large factorials in this equation imposes limitations on the attainable values of $j$, typically only a few hundred. This limitation is not only due to the memory consumption and computational complexity associated with the direct calculation of large factorials (such as $n!$ for $n>1000$), as it also stems from the numerical instability and significant precision loss. This issue is reminiscent of the longstanding problem in the numerical evaluation of the Wigner rotation matrix \cite{tajima2015analytical}, as for SCSs $|\theta,0\rangle = \exp(-i\theta J_y)|j,j\rangle=\sum_m d_{mj}^j|j,m\rangle$, where the expansion coefficients are given by the Wigner $d$-matrix.

A further examination of Eq. \eqref{scs-subeq1} shows that the SCS can be obtained directly from a unitary transformation of the polar state $|\theta=0,\phi\rangle=|j,j\rangle$. This unitary transformation is an exponential of a sparse matrix, because in the Dicke basis 
\begin{align}
    J_\pm|j,m\rangle=\sqrt{(j\mp m)(j\pm m +1)}|j,m\pm 1\rangle.
\end{align}
Exponential of sparse matrix acting on vectors can be effectively computed using Krylov subspace technique (see Ref. \cite{moler2003nineteen}). This alternative approach bypasses the problem of large factorials and enables the attainment of significantly larger value of $j$ (differing by several orders of magnitude, depending on the sparsity), compared with the previous studies. One can take an initial look at the Husimi functions of eigenstates provided in Sec. \ref{sec4.1} and more illustrations in the Appendix, where $j=2^{12}$. The Krylov subspace technique is a well-known method in numerical linear algebra and can also be applied to the numerical calculation of Wigner rotation matrices for higher spins, for comparison with other effective algorithm \cite{wang2022effective}.

\subsection{Localization of eigenstates}
One common way for defining the localization of eigenstates is via Shannon entropy. The eigenstates with odd parity, denoted as $|v_{n,-}\rangle$ can be expressed in the basis of unperturbed Hamiltonian as $v_{nm}=\langle  x_{m}|v_{n,-}\rangle$, where $|x_m\rangle$ is the eigenbasis of the unperturbed Floquet part $\exp({-i\alpha J_x})$. The Shannon entropy is defined as
\begin{align}
    H_n = -\sum_{m=1}^j |v_{nm}|^2\ln |v_{nm}|^2. 
\end{align}
It is worth noting that the basis $|x_m\rangle$ here is specifically the eigenbasis of $J_{x,-}$, where $J_x$ is restricted to odd parity. The operator $J_{x,-}$ is defined as 
\begin{align}
    J_{i,\pm}=P_{\pm}J_iP_{\pm}, \ P_{\pm}=(1\pm R_x)/2,\quad i=x,y,z.
\end{align}
Here $P_\pm$ are projection operators associated with parity.
For the even parity, the Shannon entropy can be defined in a same way. We have verified that the statistics of $H_n$ is nearly identical for both parities. This is also referred to as the eigenvector statistics, which is obviously basis-dependent. 

The Shannon ELM is defined as $L_n= e^{H_n}/N_{odd}$, with the dimension of odd-parity subspace $N_{odd}=j$, and we denote the mean Shannon ELM as $\langle L\rangle$. In Appendix \ref{app-proof1}, we have proven that for $j\gg1$, the Shannon ELM is the same whether within the parity subspace or not, and the Shannon entropy is almost the same if $H_n\gg1$. However, this is not the case for level statistics, where the Hilbert space must be split according to parity, because energy levels with different parity may not exhibit level repulsion.

Apart from this basis-dependent method,  another approach relies on the Wehrl entropy of the eigenstates, based on the overcomplete basis of coherent state, uniquely determined by the classical phase space. It is defined as 
\begin{align}
    \label{eq:wehrl}
    S_n =-\frac{2j+1}{4\pi}\int \sin\theta d\theta d\phi\  Q_n(\theta,\phi)\ln Q_n(\theta,\phi).
\end{align}
The Wehrl ELM is $\mathcal{L}_n=e^{S_n}/N$, with $N=2j+1$ and  $\langle {\mathcal{L}}\rangle $ denotes the averaging. It's worth noting that Wehrl conjectured that the minimum of Wehrl entropy is attained for coherent state \cite{wehrl1979relation}. Lieb originally proved the conjecture \cite{lieb1978proof}, later extended it to SU(2) and SU(N) coherent states \cite{lieb1991quantum,lieb2014proof,lieb2016proof}. However, the problem of the uniqueness of minimizers in some cases is still open \cite{schupp2022wehrl}.

One may anticipate that a random eigenstate will tend to the uniform distribution $|v_{nk}|^2= 1/N_{odd}$ and $Q_n(\theta,\phi)=1/N$, resulting in the maximum value of both entropy $H_n=\ln N_{odd}$ and $S_n=\ln N$, as well as ELMs $L_n = \mathcal{L}_n = 1$. For a random pure state $|\mathcal{R}\rangle$ from COE/GOE \cite{wootters1990random,jones1990entropy}, it was proven that mean Shannon entropy $\langle H\rangle_\mathcal{R} =\Psi(N/2+1)-\Psi(3/2)\sim \ln N+\gamma_e+\ln2-2\approx \ln N -0.7296$, with $\Psi(x)$ the digamma function and $\gamma_e$ denotes the Euler constant, where the approximation is for the asymptotic limit $N\to \infty$. Additionally, the mean Shannon/Wehrl entropy of a random pure state from CUE is $\langle S\rangle_\mathcal{R} =\Psi(N+1)-\Psi(2)\sim \ln N+\gamma_e-1\approx \ln N-0.4228$ \cite{gnutzmann2001renyi}. The random pure state statistics then gives $\langle L\rangle_\mathcal{R} \approx0.482$, and $\langle \mathcal{L}\rangle_\mathcal{R} \approx 0.655$ for large $N$.

\begin{figure}
    \includegraphics[width=0.95\linewidth]{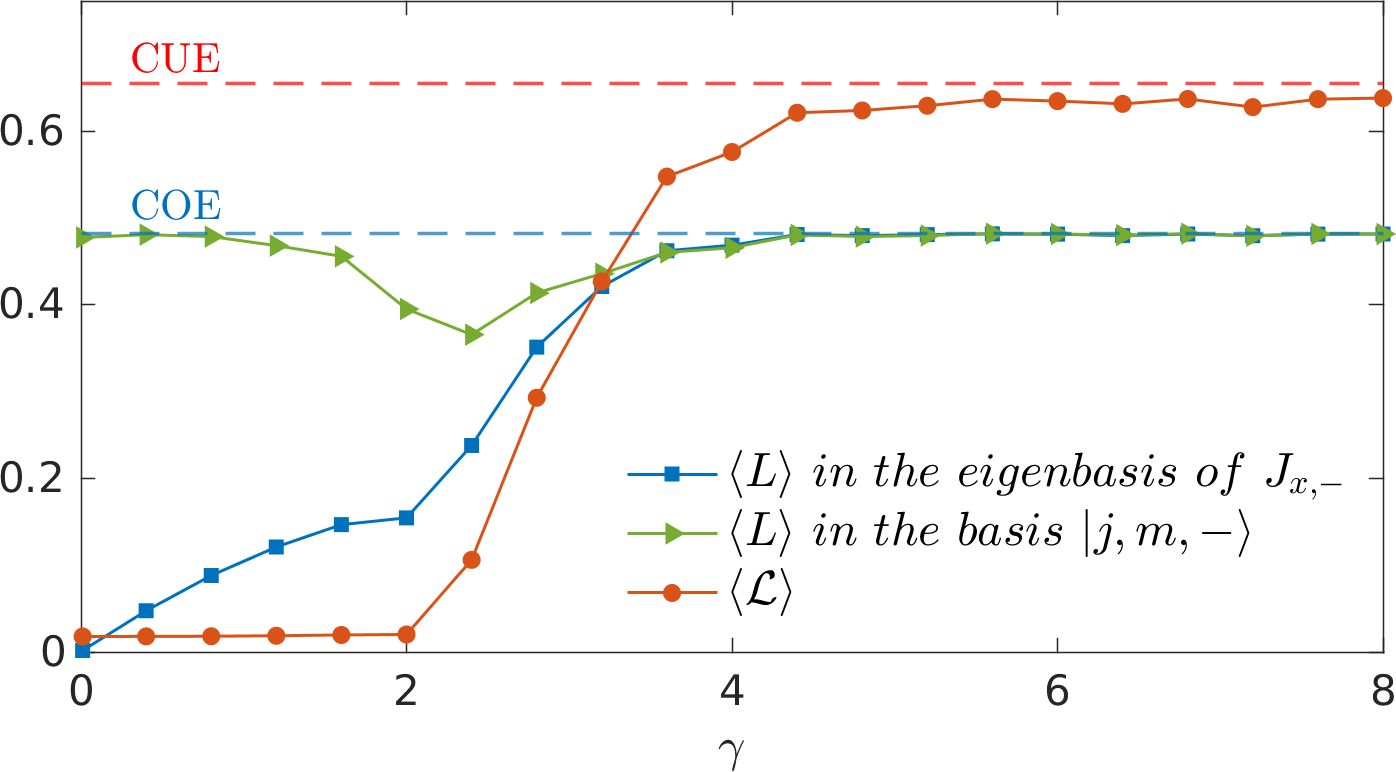}
    \caption{Mean Shannon ELM $\langle L\rangle$ calculated for the odd parity, in the eigenbasis of $J_{x,-}$ and the basis $|j,m,-\rangle$ for odd parity defined in Eq. \eqref{eq:parity-basis} and mean Wehrl ELM $\langle \mathcal{L}\rangle$, versus $\gamma$ for QKT with $j=2^{12}$ and $\alpha=11\pi/19$. The dashed horizontal lines indicate the mean Shannon ELM from COE statistics and Wehrl/Shannon ELM of CUE.} 
    \label{fig:LocalLength}
\end{figure}

In Fig. \ref{fig:LocalLength}, we study the mean Shannon ELM $\langle L\rangle$ in the eigenbasis of $J_{x,-}$ and the basis $|j,m,-\rangle$ for odd parity defined in Eq. \eqref{eq:parity-basis}, for eigenstates with odd parity (we found for the even parity, it is the same), as well as the mean Wehrl ELM $\langle \mathcal{L}\rangle$.  It shows that the Shannon ELM in  the basis of $J_{x,-}$ saturates to the COE value $\langle L\rangle_R $ for $\gamma \gtrsim 4.4$, and before that, it exhibits two different types of growth: for $\gamma > 2$, it grows faster than for $\gamma \leq 2$. While $\langle L\rangle$ in $|j,m,-\rangle$ basis also eventually reaches the COE value $\langle L\rangle_R $ for $\gamma \gtrsim 4.4$, but the growth pattern leading up to saturation is markedly different and does not follow the classical transition route to chaos (as shown in Fig. \ref{fig:FractionRatio}(a), and see Appendix \ref{app-proof1} for further discussions). This difference is clearly due to the basis-dependence of Shannon entropy. The Wehrl ELM is calculated by using a grid of $N_p=200\times 400$ points in the $(\theta,\phi)$ space, with each point holding a coherent state, where all the Husimi functions are normalized according to the discretization of Eq. \eqref{eq:norm2}, as
\begin{align}
 \label{eq:norm-dis}
\frac{(2j+1)\pi}{2N_p}\sum_{ij}Q_n(\theta_i,\phi_j)\sin\theta_i=1.
\end{align}
The transition of mean Wehrl ELM versus $\gamma$ follows exactly the classical route. For $\gamma \leq 2$, it maintains a value equal to $\langle \mathcal{L}\rangle_{J_x}\sim N^{-1/2}$, representing the mean Wehrl ELM of the operator $J_x$ in the basis of unperturbed Hamiltonian (see proof in Appendix \ref{app-proof2}), which is close to 0 for large $N$. It then increases until it saturates for $\gamma \gtrsim 4.4$, to a value close to $\langle \mathcal{L}\rangle_R$ from CUE. 

\begin{figure}
    \includegraphics[width=0.9\linewidth]{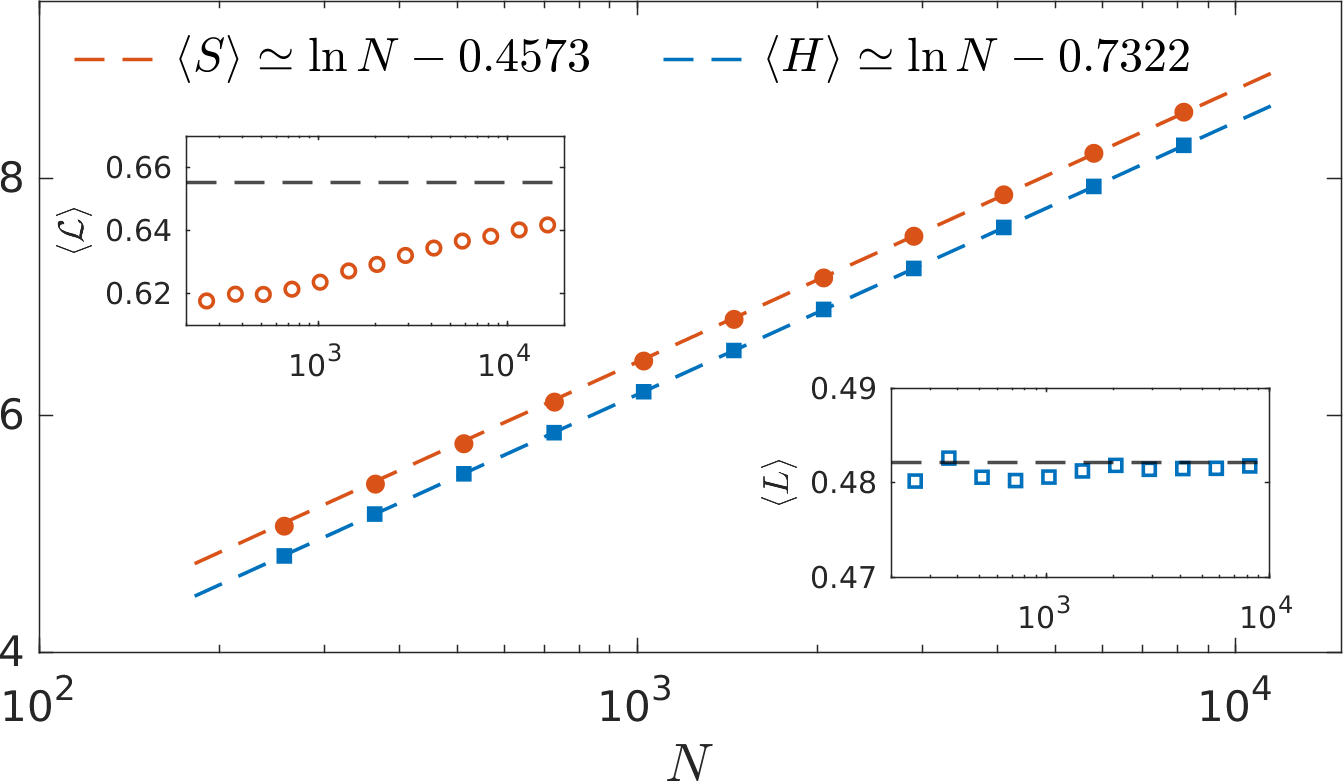}
    \caption{Mean Shannon entropy $\langle H\rangle$ and mean Wehrl entropy $\langle S \rangle$, versus $N$ for QKT, with $\gamma=8$ and $\alpha=11\pi/19$, where for the former $N=N_{odd}=j$ and the latter $N=2j+1$. 
    Two dashed lines are from best fitting. Two insets show two mean ELMs versus $N$, respectively, where the  dashed lines indicate $\langle \mathcal{L}\rangle_\mathcal{R}$ from CUE (top left) and $\langle L\rangle_\mathcal{R}$ from eigenvector statistics of COE (bottom right).}
    \label{fig:qkt-rmt}
\end{figure}

Numerically, a larger $j$ requires more grid points, because the number of Planck cells is $N_c \sim \hbar_{eff}^{-1} = j$. In this work, for the following calculations, we set $N_p=200\times 400 \gg j$, for the subsequent calculations to ensure the convergence of all numerical results (see details about the numerical convergence in Appendix \ref{app-proof2}). 
Our findings indicate that the mean Wehrl entropy of fully chaotic QKT is slightly smaller than that $\langle S\rangle_\mathcal{R}$ from CUE, and it was conjectured in Ref. \cite{gnutzmann2001renyi} that the difference between the mean Wehrl entropy of COE and CUE vanishes in the asymptotic limit $N\to \infty$. However, both a numerical verification at larger $N$ and a formal mathematical proof of this conjecture are still lacking. In this work, using the Krylov subspace technique to generate SCSs, we study the mean Shannon entropy $\langle H\rangle$ and Wehrl entropy $\langle S\rangle$, of the QKT at $\gamma=8$ where the system is fully chaotic, as shown in Fig. \ref{fig:qkt-rmt}. The eigenvector statistic reveals that $\langle H\rangle\simeq \ln N-0.7322$, is nearly equal to $\langle H\rangle_\mathcal{R}$ of COE. For Shannon ELM, $\langle L\rangle \simeq \langle L\rangle_\mathcal{R}$, as shown in the inset (bottom right). On the other hand, the mean Wehrl entropy $\langle S\rangle\simeq \ln N - 0.4573$,  is close to $\langle S\rangle_\mathcal{R}$ of CUE, while the inset in the top left shows that $\langle \mathcal{L}\rangle$ approaches the CUE value $\langle \mathcal{L}\rangle_\mathcal{R}$ with increasing system size. 

In fully chaotic QKT exhibiting COE spectral statistics, the distinct ergodic properties of coherent states in the eigenbasis $|v_n\rangle$ and the eigenvectors in the angular momentum basis account for the difference between the CUE result for the mean Wehrl entropy and the COE result for the mean Shannon entropy. Consider chaotic systems without any antiunitary symmetries, described by $N\times N$ CUE random matrices, or systems with one antiunitary symmetry, such as time-reversal, described by COE random matrices. As $N\to \infty$, the components of eigenvector $|\psi_n\rangle$ of random matrices, denoted by $x_i=|\langle i|\psi_n\rangle|^2$, follow the $\chi_\nu^2$ distribution \cite{zyczkowski1990indicators}
\begin{align}
    P_\nu(x)=\left(\frac{\nu}{2\langle x\rangle}\right)^{\nu/2}\frac{1}{\Gamma(\nu/2)}x^{\nu/2-1}\exp\left(-\frac{\nu x}{2\langle x\rangle}\right),
\end{align}
where $\langle x\rangle$ is the mean value of $\{x_i\}$, equal to $1/N$. For the COE, $\nu=1$, $P_\nu(x)$ is the well-known Porter-Thomas distribution \cite{porter1956fluctuations}. For the CUE,  $\nu=2$, and for symplectic ensemble, $\nu=4$. This distribution represents the sum of squares $x=\sum_{i=1}^\nu y_i^2$ if each $y_i$ is an independent random variable with a Gaussian probability distribution, with the mean equal to zero and the variance equal to $\sqrt{\langle x\rangle}/\nu$.  For large $N$, there are a significant number of small components, so it is more effective to use the distribution of $\{\ln x_i\}$ instead of $\{x_i\}$, given by \cite{wang2021multifractality}
\begin{align}
   P_\nu(&\ln x)=\frac{P_\nu(x)}{d\ln x/dx}\nonumber \\
    &=\left(\frac{\nu}{2\langle x\rangle}\right)^{\nu/2}\frac{1}{\Gamma(\nu/2)}x^{\nu/2}\exp\left(-\frac{\nu x}{2\langle x\rangle}\right).
\end{align}
The relation $P_\nu(\ln x)=xP_\nu(x)$ implies that $P_\nu(\ln x)$ has the maximal value at $x=\langle x\rangle=1/N$.

\begin{figure}
    \includegraphics[width=1\linewidth]{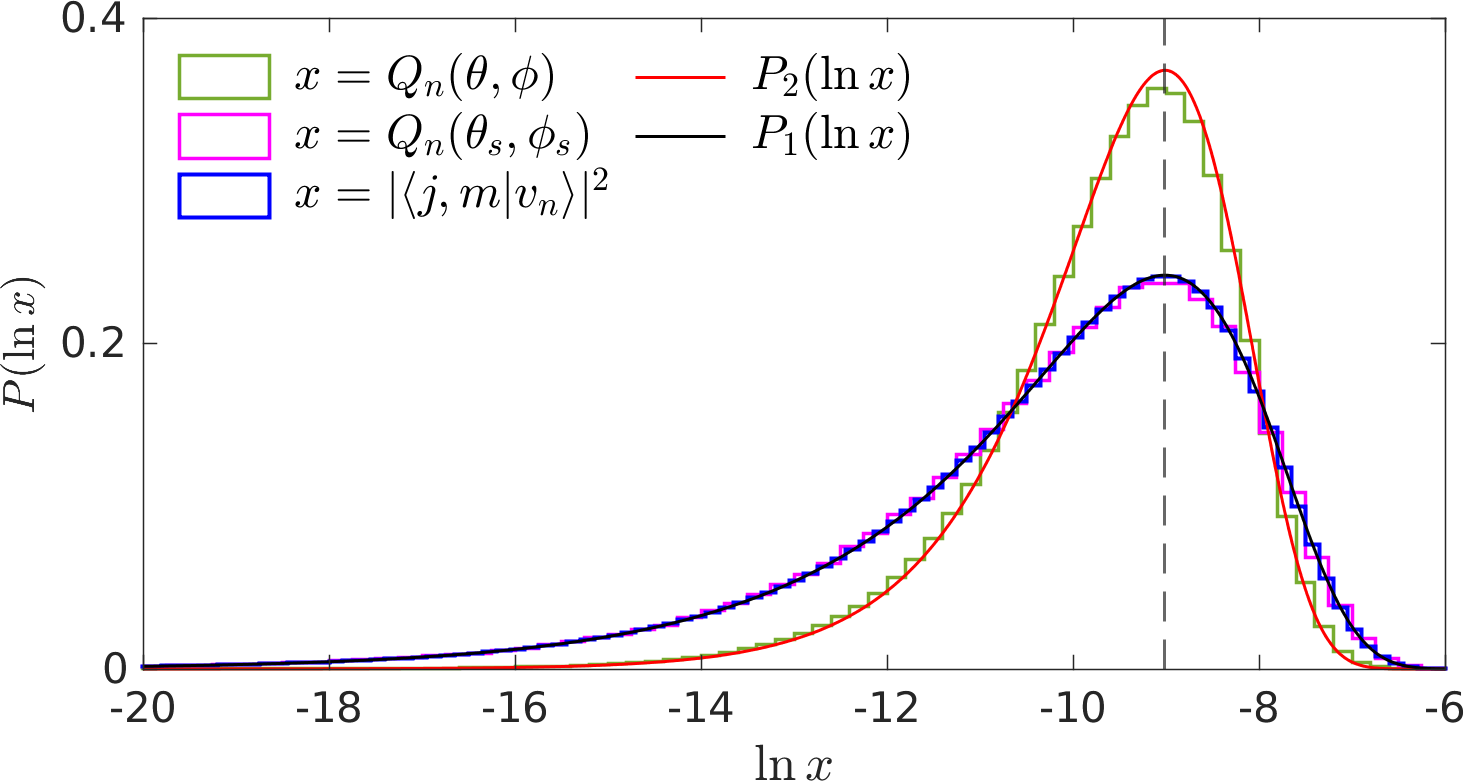}
    \caption{Histograms of $P(\ln x)$ for QKT with $\gamma=8$, the vertical dash line denotes $x=\langle x\rangle =1/N$ with $N=2j+1$, the solid lines represent the $\chi_\nu^2$ distribution. $P(\ln Q_n (\theta,\phi))$ has been calculated from $8\times 10^4$ coherent states held on $200\times 400$ grid points in phase space, and $P(\ln Q_n (\theta_s,\phi_s))$ from $300$ coherent states along the symmetry lines, both in the basis of $2j+1$ eigenvectors. Parameters: $j=2^{12}$, $\alpha=11\pi/19$.}
    \label{fig:eigenvDist}
\end{figure}

The expansion coefficients of some fixed vector in the eigenbasis of random matrices will have the same statistical properties.  Specifically, the components $|\langle \alpha|\psi_n\rangle|^2$ of an arbitrary coherent state $|\alpha\rangle$ in the eigenbasis $|\psi_n\rangle$ follow a $\chi_\nu^2$ distribution. In a fully chaotic QKT, the components $Q_n(\theta,\phi)$ of the SCS $|\theta,\phi\rangle$ in the eigenbasis $|v_n\rangle$ of the Floquet operator should conform to this distribution, $\nu=2$. Similarly, the expansion coefficients $|\langle j,m|v_n\rangle|^2$ of eigenvectors in the $|j,m\rangle$ basis should follow the distribution, $\nu=1$. In Fig. \ref{fig:eigenvDist} we show $P(\ln x)$ for both types of expansion coefficients in QKT with $\gamma=8$, which is fully chaotic. It demonstrates that for all SCSs located on the unit sphere, $P(\ln Q_n)$ is closely matches $P_2(\ln x)$, while $P(\ln |\langle j,m|v_n\rangle|^2)$ is exactly described by $P_1(\ln x)$. An intuitive explanation of this difference is that the expansion coefficients of SCS in the eigenbasis are generally complex numbers, where the real part and the imaginary part are independent Gaussian random numbers. 

A more physical argument proposed by Zyczkowski \cite{zyczkowski2001localization} suggests that there exists a symmetry line in the phase space and the coherent state located along this line exhibit a \(\chi_1^2\) distribution of COE, while those off the line follow a \(\chi_2^2\) distribution of CUE.  Importantly, this single symmetry line has measure zero within the compact phase space. However, instead of a single symmetry line in the phase space, there exists a family of symmetry curves in the classical phase space. These curves are defined as the set of points $\mathbf{x}\in \mathbb{S}^2$  that satisfy the condition $\mathcal{I}(\mathbf{x})=\mathbf{x}$, where $\mathcal{I}=T_1^c,T_2^c,T_1^cF_c,F_cT_1^c,\cdots$ \cite{haake1987classical,munoz2021nonlinear}. For instance, the symmetry lines of $T_1^c$ and $T_2^c$ are given by 
\begin{align}
    y\sin\alpha+z\cos\alpha=\pm z,
\end{align}
are great circles on the unit sphere, a parametrization of these two lines on $\mathbb{S}^2$ is
\begin{align}
    \tan\theta_s\sin\phi_s=a_s, \quad a_s=\frac{-\cos\alpha\pm1}{\sin\alpha}.
\end{align}
The symmetry lines for higher-order involutions tend to have more complicated shapes. 

In Fig. \ref{fig:eigenvDist} we show that for SCSs $|\theta_s,\phi_s\rangle$ located on two great circles, the distribution of the expansion coefficients $P(Q_n(\theta_s,\phi_s))$ exactly follows $\chi_1^2$ distribution. 
We can formally prove that the family of symmetry lines $\{l_n\}$, has zero measure in the phase space. In the compact space $\mathbb{S}^2$, by defining the Lebesgue outer measure $m^*$ for each $l_n$, it is evident each line on $\mathbb{S}^2$ has $m^*(l_n)=0$. Considering all the involutions in $\mathcal{I}$, one can establish a bijection of each involution and the natural number, proving that $\{l_n\}$ is countable with $ n\in\mathbb{N}$. From the weaker property of countable subadditivity \cite{axler2020measure,tao2011introduction}, we see
\begin{align}
    m^*(\bigcup_{n\in \mathbb{N}} l_n)\le \sum_{n\in\mathbb{N}}m^*(l_n)=0,
\end{align}
meaning that the symmetry lines are of zero measure. Assuming the distribution of $Q_n(\theta,\phi)$ is independent of $\theta$ and $\phi$, for the mean Wehrl entropy,  we have
\begin{align}
    \langle S\rangle &=-\frac{2j+1}{4\pi}\int  Q_n\ln Q_nP(Q_n)\sin\theta d\theta d\phi dQ_n\nonumber \\
    &\simeq -\frac{2j+1}{4\pi}\int \sin\theta d\theta d\phi \int Q_n\ln Q_nP_2(Q_n)dQ_n\nonumber \\
    &=-(2j+1)\int_0^\infty Q_n\ln Q_nP_2(Q_n)dQ_n\nonumber \\
    &=\ln N-\Psi(2) ,
\end{align}
where $N=2j+1$,  which is exactly $\langle S\rangle_{\mathcal{R}}$ of CUE in the asymptotic limit.

\begin{figure*}
    \includegraphics[width=0.75\linewidth]{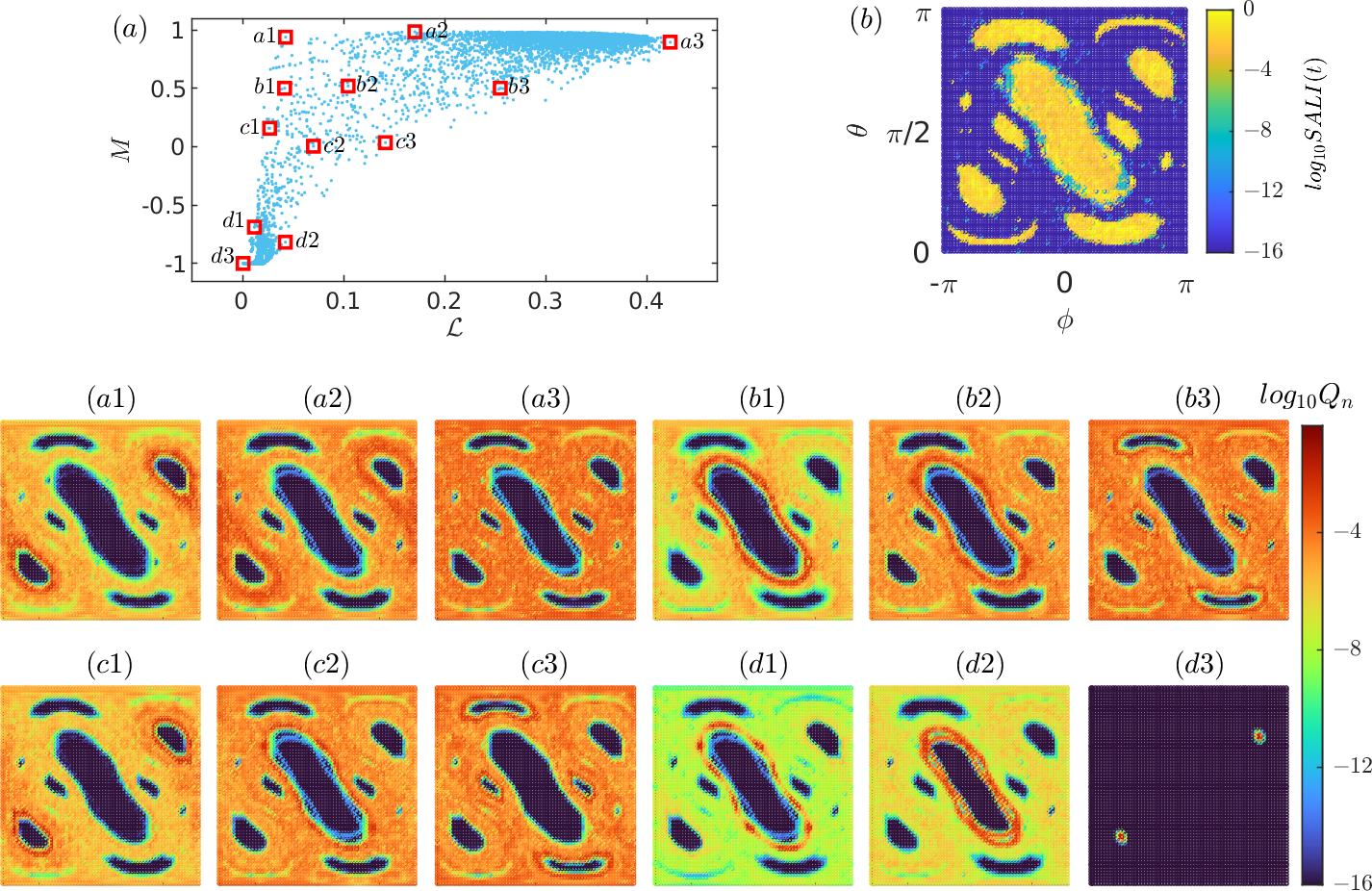}
    \caption{(a) A joint distribution  $(\mathcal{L},M)$ for 8193 eigenstates of QKT Floquet operator, with $\gamma=2.6$, $\alpha=11\pi/19$ and $j=2^{12}$. States cluster at the overlap index extremes—regular at one end, chaotic at the other—with mixed states sparsely in between. Husimi functions of eigenstates isolated by boxes at different positions are plotted from panel (a1) to (d3), the darkest blues show the area where $Q_n< 10^{-16}$. (b) Logarithmic values of SALI for the corresponding CKT, on the parametrized phase space $(\theta,\phi)$ after 300 kicks.}
    \label{fig:husimi}
\end{figure*}

\section{Phase-space overlap index}
\label{sec4}
The Husimi functions of eigenstates in quantum maps \cite{bogomolny1992semiclassical,prosen1996quantum},  are precise analog of the classical Poincar\'e section. From this correspondence, the Husimi functions of eigenstates can be used for the identification of regular, mixed, and chaotic eigenstates, by the criterion of overlap with the classical Poincar\'e section, where one can employ the SALI method to identify whether an initial condition belongs to the classical chaotic or regular regions. As it was introduced and implemented in previous works  \cite{batistic2013dynamical,batistic2013intermediate,batistic2013quantum}, for eigenstate $|v_n\rangle$, the overlap can be quantified by the overlap index $M_n$, defined as
\begin{align}
    M_n=\frac{2j+1}{4\pi}\int \sin\theta d\theta d\phi\ f(\chi_c) Q_n (\theta,\phi)  ,  
\end{align}
where \( f(\chi_c) = 1 \) if \( \chi_c = 1 \), and \( f(\chi_c) = -1 \) if \( \chi_c = 0 \). Here, \( \chi_c(\theta,\phi) \) denotes the characteristic function of the chaotic component, as defined in Eq. \eqref{eq:mu-c}. 

According to PUSC in mixed-type systems, for Husimi functions of eigenstates,  $M$ should be either $+1$ or $-1$ in the strict semiclassical limit, corresponding to chaotic or regular eigenstates, respectively. However, since the semiclassical limit is not easily reached in practice, $M$ actually varies between $-1$ and $+1$. Upon approaching the semiclassical limit, a reduction in the fraction of mixed eigenstates with $|M|<1$ would occur, as evidenced first in the billiard system \cite{lozej2022phenomenology}, particularly exhibiting a power-law decay with increasing energy, i.e. decreasing effective Planck constant. In the following, we analyze quantum eigenstates of Floquet operator in mixed-type QKT by studying the joint distribution of Wehrl ELM and overlap index. Then, we extend the analysis of the power-law decay of the fraction of mixed eigenstates as system size increases, as conducted in  Ref. \cite{wang2023power} from a few hundred, now up to $j=2^{13}$. For numerical calculation of Husimi functions, we use a grid of $N_p=200\times 400$ points in the $(\theta,\phi)$ space, with each point holding a coherent state.

\begin{figure*}
    \includegraphics[width=0.75\linewidth]{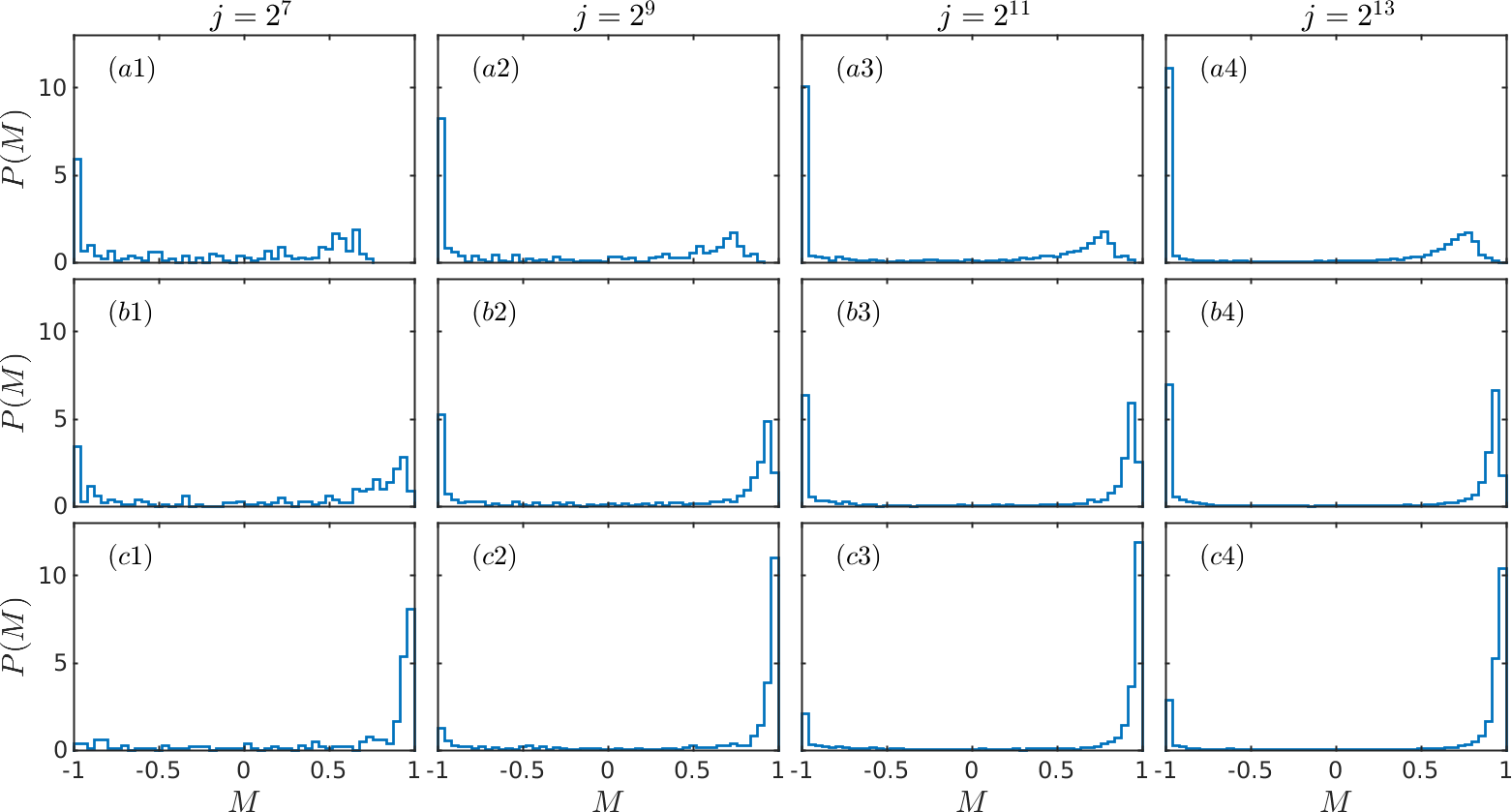}
    \caption{Histograms of $P(M)$ for increasing system size, from left to right. In panels (a1)-(a4), the kicking strength is set to $\gamma=2.3$. In panels (b1)-(b4), it is $\gamma=2.6$, and in panels (c1)-(c4), it is $\gamma=3$.}
    \label{fig:MixedDist}
\end{figure*}

\subsection{The overlap index and Wehrl ELM}
\label{sec4.1}
From the definition of Wehrl ELM following Eq. \eqref{eq:wehrl},  regular eigenstates characterized by an overlap index of $M=-1$ and condensing on invariant tori are localized, yielding lower values of $\mathcal{L}$. In contrast, chaotic eigenstates with $M=+1$ can be either strongly localized or extended. Previous studies have numerically shown that the probability distribution $P(\mathcal{L})$ of chaotic eigenstates indeed follows a beta distribution \cite{batistic2019statistical,batistic2020distribution,yan2024chaos}. As the system approaches the semiclassical limit, it tends towards a delta distribution $\delta(\mathcal{L}-\langle \mathcal L\rangle_\mathcal{R})$. It is important to note that in mixed-type systems, to verify this distribution, the Wehrl ELM of the chaotic eigenstates should be further renormalized by the relative area $\mu_c$ of the chaotic component in the phase space.  In addition to strictly regular and chaotic states, there would exist mixed eigenstates with $|M|<1$ in mixed-type systems, exhibiting varying degrees of localization. This is illustrated in Fig. \ref{fig:husimi}(a), plotting the joint distribution $(\mathcal{L},M)$ for QKT at $\gamma=2.6$, alongside selected representative Husimi functions of eigenstates from different regions of the parameter space, as shown in Fig. \ref{fig:husimi}(a1)-(d3). Moreover, Fig. \ref{fig:husimi}(b) presents the SALI plot from the corresponding classical dynamics at $\gamma=2.6$ in phase space, for comparison. At the border of the regular islands and chaotic sea, visible stickiness are denoted by intermediate colors.

Fig. \ref{fig:husimi}(a) shows that states predominantly cluster at both ends of the overlap index, one end representing regular states and the other chaotic states, while mixed states sparsely distribute in between. In comparison, for smaller kicking strengths, such as $\gamma = 2.3$, shown in Fig. \ref{fig:husimi1}(a) (see Appendix \ref{app-gallery} the gallery of states), the clustering is less pronounced. At $\gamma=3$ (Fig. \ref{fig:husimi2}(a)), the clustering becomes more evident, as larger values of $\gamma$ correspond to larger $\mu_c$ -- the fraction of chaotic components, and a simpler structure of classical phase space, resulting in a smaller region between the chaotic sea and regular islands where stickiness occurs. Examining Husimi functions of the chaotic eigenstates plotted in Fig. \ref{fig:husimi}(a1)-(a3), alongside the corresponding SALI plot, a reduction in the area of the darkest blue regions (with $Q_n\le 10^{-16}$) belonging to regular islands can be observed. The most extended state shown in Fig. \ref{fig:husimi}(a3), infiltrates into the regular islands due to quantum tunneling, and exhibits a minor overlap with the outer tori of the regular islands. The state with the maximum overlap index, as plotted in Fig. \ref{fig:husimi}(a1), is strongly localized around and avoids the outer tori of the regular region. Similar observation extends to another layer of overlap index $M \simeq 0.5$ for states illustrated in Fig. \ref{fig:husimi}(b1)-(b3), and to $M \simeq 0$ in Fig. \ref{fig:husimi}(c1)-(c3), except that the most localized ones now live in the vicinity of the regular islands, with smaller contribution from the chaotic sea.  Fig. \ref{fig:husimi}(d1)-(d3) illustrate states from the bottom layer of overlap index. Particularly, Fig. \ref{fig:husimi}(d1) shows a state predominantly localized at a chain of regular islands, which exhibits tunneling between these regular islands. In Fig. \ref{fig:husimi}(d2), the state also resides in the vicinity of the regular island, with visible stickiness surrounding it, appearing more extended. The most localized state shown in Fig. \ref{fig:husimi}(d3) is located exactly at the fixed points.

In summary, the joint distribution $(\mathcal{L},M)$ offers a comprehensive overview of mixed-type QKT eigenstates. The study of Husimi functions shows that, mixed states predominantly occupy chaotic regions with $M > 0$, flooding into regular areas, while states concentrate in regular islands with $M < 0$, displaying tunneling between them. Furthermore, among states with a smaller overlap index, the Wehrl ELM shows a narrower range.

\subsection{The fraction of mixed eigenstates}
\label{4.2}

In the semiclassical limit, it is expected that mixed states gradually disappear in accordance with the Berry-Robnik picture and PUSC. This is supported by the histograms depicted in Fig. \ref{fig:MixedDist}, which illustrate the distribution $P(M)$ of the overlap index $M$ as the system size $j$ increases, at different kicking strengths. As the semiclassical limit is approached with increasing $j$, we observe that: (i) Increasing values of $j$ lead to a more pronounced alignment of eigenstates with either regular or chaotic clusters. Specifically, at $\gamma=2.3$, the kicking strength closer to the integrability breaking at $\gamma\simeq 2$, the chaotic cluster centers around $M\simeq 0.8$, correlating with a much more complex phase space structure (see Fig. \ref{fig:husimi1} in Appendix \ref{app-gallery}) and the possible classical partial transport barriers \cite{mackay1984transport,geisel1986kolmogorov,bohigas1990dynamical,michler2012universal,stober2024quantum,meiss2015thirty}. (ii) Fluctuations among intermediate values of $M$ decrease as the system size $j$ increases. (iii) Increasing $\gamma$ enhances chaos and $\mu_c$, reflected in $P(M)$ with a higher peak at $M = +1$ and a lower peak at $M=-1$.

\begin{figure}
    \includegraphics[width=0.9\linewidth]{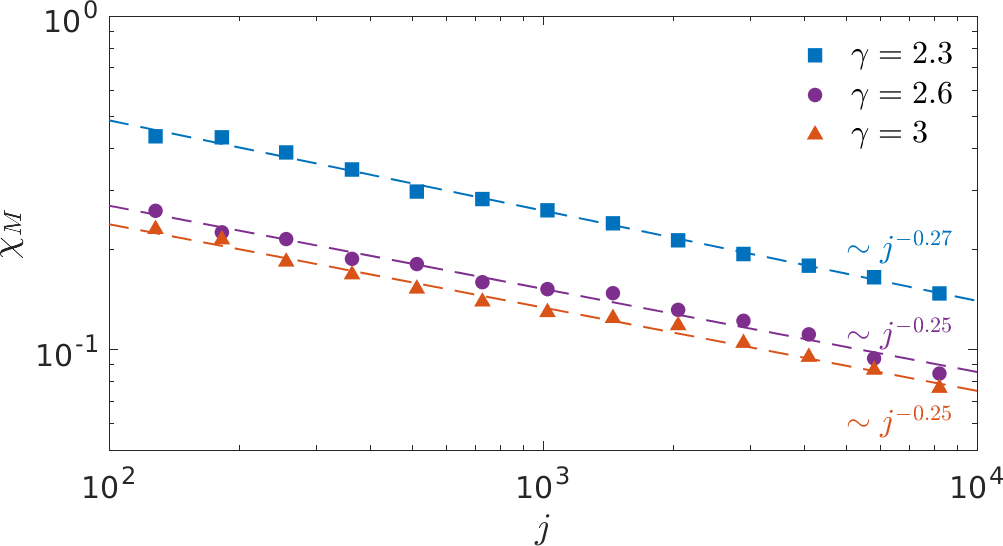}
    \caption{Decay of the fraction of mixed eigenstates $\chi_M$ with respect to $j$, at three kicking strengths $\gamma=2.3$ (squares), $\gamma=2.6$ (circles) and $\gamma=3$ (triangles). The mixed eigenstates criteria here is $M\in[-0.9,0.5]$. The dashed lines show the power law decay of $\chi_M \sim j^{-\zeta}$. Here, the maximal $j=2^{13}$.}
    \label{fig:MixedDecay1}
\end{figure}

Following these observations, to quantify the decay of the fraction of mixed states with respect to the increasing system size $j$, we define
\begin{align}
    \chi_M = N(M\in [M_0,M_1])/N,
\end{align}
as the fraction of eigenstates within an interval of the overlap index $M_0\le M\le M_1$, given the total number of states $N=2j+1$. In Fig. \ref{fig:MixedDecay1}, we show the case where $M\in[-0.9,0.5]$ and plot $\chi_M$ as a function of $j$ for different $\gamma$. This behavior is well described (fitted) by a power-law decay, $\chi_M\sim j^{-\zeta}$, with the power exponent $\zeta$ being very similar for different $\gamma$. Here, the range of $j$ we considered is expanded by two orders of magnitude compared to the previous study \cite{wang2023power}, which only examined $j$ up to 400, resulting in substantially improved fitting.
 
\begin{figure}
    \includegraphics[width=0.9\linewidth]{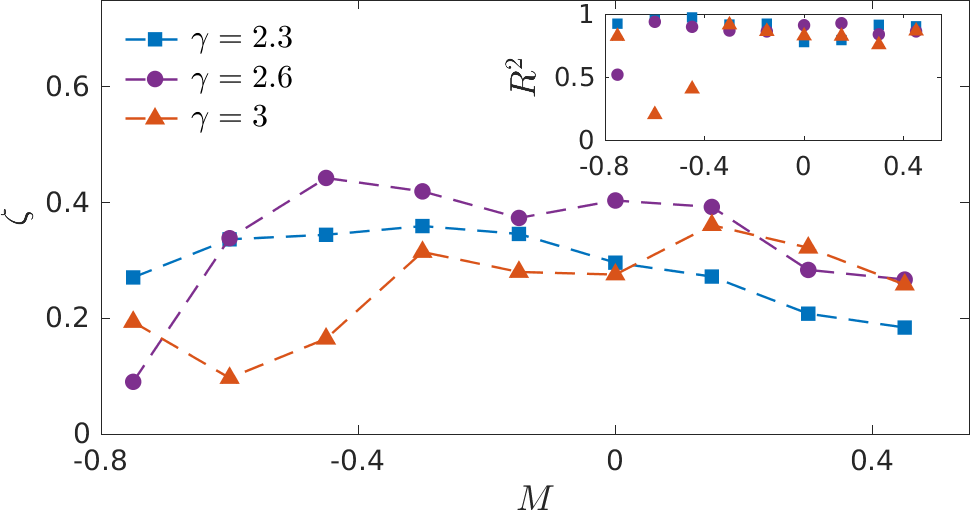}
    \caption{Power-law decay exponent $\zeta$ for $M\in[M-\Delta M/2,M+\Delta M/2]$, with $\Delta M = 0.3$, across different intervals. The inset shows the goodness of all best-fittings, indicated by the coefficient of determination $R^2$.}
    \label{fig:MixedDecay2}
\end{figure}

The decay exponent $\zeta$ certainly depends on the interval we consider for $M$. The interval should not be too small, lacking sufficient states for good statistics, or too large to show little variation across different intervals. In Fig. \ref{fig:MixedDecay2}, we illustrate the variation of the decay exponent when considering small intervals $M\in[M-\Delta M/2,M+\Delta M/2]$, with $\Delta M = 0.3$, for different $\gamma$. The inset displays the goodness of the best-fitting, indicated by $R^2$, the coefficient of determination. Disregarding a few imperfect fittings, the data suggests a decay exponent range of $0.2 < \zeta < 0.45$.  Given the similar power-law decay observed in the fraction of mixed eigenstates in both billiards \cite{lozej2022phenomenology} and quantum H\'enon-Heiles system \cite{yan2024chaos}, along with the Dicke model \cite{wang2024mixed}, we conjecture that this behavior is a universal property of mixed states, as the system approaches the semiclassical limit. A more profound open question is  whether there exists an upper bound of the decay exponent.

We emphasize that in Ref. \cite{ketzmerick2000new}, the hierarchical states were introduced as a new class of eigenstates in mixed-type systems, distinct from the regular and chaotic ones. These states, observed in the quantum kicked rotor with much simpler phase space structure, predominantly live near the regular islands, with only a minor contribution in the main part of the chaotic sea. They also demonstrate a power-law decay $\hbar^\alpha\ (\alpha>0)$, attributed to the transport properties associated with stickiness between regular islands and the chaotic sea. From this definition, the hierarchical states mostly are highly localized states as shown in Fig. \ref{fig:husimi}(b1), (c1), (d1) (or the same labeled states in Fig. \ref{fig:husimi1}-\ref{fig:husimi2}). While the overlap index captures all the mixed states across different degrees of localization, it is influenced also by the flooding of chaotic sea to regular islands or the tunneling between regular islands,
not directly related to the effects of classical stickiness. So it would be interesting to study the behavior of mixed states in systems without classical stickiness, like certain lemon billiards \cite{lozej2021classical,lozej2021effects}.

\section{Conclusions}
\label{sec5}
In this work, we employed the Krylov subspace technique to generate spin coherent states, thereby extending the investigation of the statistics of Husimi functions of eigenstates in the kicked top, a prototype model for studying quantum chaos. Notably, we explored system sizes up to $j=2^{13}$, a significant advancement compared to previous studies of Husimi functions limited to $j$ in the range of a few hundred. We examined the transition to chaos in both classical and quantum case. All the indicators show a good quantum-classical correspondence, specifically the mean Wehrl entropy localization measure in the statistics of eigenstates. With this new spin coherent states generating method, we also demonstrated that the mean Wehrl entropy localization measure of the fully chaotic quantum kicked top, which follows the COE spectral statistics, approaches the CUE value as the system size $j$ increases, which is surprising \cite{gnutzmann2001renyi}. This indicates that the statistics followed by the eigenvalues and eigenvectors do not necessarily follow the same ensemble. We proved that coherent states located on symmetry lines of the classical phase space exhibit COE eigenvector statistics but have a zero measure, while other coherent states display eigenvector statistics consistent with CUE.

More importantly, in the case of a mixed-type kicked top where regular islands and chaotic sea coexist in the classical compact phase space, we examined mixed eigenstates using a joint distribution of the overlap index $M$ and Wehrl localization measure $\mathcal{L}$. \(M=+1\) indicates chaotic states, while \(M=-1\) indicates regular ones. We extended the study of mixed eigenstates with $|M|<1$ in our previous investigation \cite{wang2023power} from $j=400$ to $j=2^{13}$, confirming the power-law decay of the fraction of mixed states across two orders of magnitude in $j$. This provides supporting evidence for the PUSC and the Berry-Robnik picture in the semiclassical limit. Based on this result from quantum kicked top, and similar results from other models \cite{lozej2022phenomenology,yan2024chaos,wang2024mixed}, we conjecture that the power-law decay is a universal property for mixed eigenstates. This hypothesis, particularly the connection with classical stickiness or partial barriers, merits further theoretical scrutiny.

\section{Acknowledgement}
This work was supported by the Slovenian Research and Innovation Agency (ARIS) under the grant J1-4387. 

\appendix

\section{Parity and eigenvector statistics}
\label{app-proof1}
There are two equivalent ways to split the Hilbert space of the quantum kicked top, as defined in Eq. \eqref{eq:KT-Hamiltonian}, into even- and odd-parity subspace. The first method involves rewriting the Hamiltonian matrix in a new basis
\begin{align}
    |j,m,\pm\rangle=(|j,m\rangle\pm|j,-m\rangle)/\sqrt{2},
\end{align}
where $|j,0,+\rangle$ is not normalized. On this basis, the odd-parity subspace is spanned by the states $|j,m,-\rangle$ for $1 \le m \le j$, while the even-parity subspace is spanned by the states $|j,m,+\rangle$ for $0 \le m \le j$. The second method is formally as follows:
\begin{align}
F|v_n\rangle = e^{iv_n}|v_n\rangle, \quad R_x|v_n\rangle = \pm|v_n\rangle,
\end{align}
where $|v_{n,-}\rangle$ denotes the eigenvector $|v_n\rangle$ satisfying $R_x|v_n\rangle = -|v_n\rangle$, and $|v_{n,+}\rangle$ denotes the eigenvector satisfying $R_x|v_n\rangle = +|v_n\rangle$. This equivalence implies $|v_{n,\pm}\rangle$ are also eigenvectors of $F^\pm$ defined in Eq. \eqref{eq:floquet-parity}. Expand the eigenvectors in the new basis 
\begin{align}
    |v_{n,\pm}\rangle =\sum_m c_{nm,\pm}|j,m,\pm\rangle.
\end{align}
If restricted $|v_{n,-}\rangle$ in the odd-parity subspace, the Shannon entropy is 
\begin{align}
    H_n = -\sum_{m=1}^{j} |c_{nm,-}|^2\ln |c_{nm,-}|^2.
\end{align} 
Without restriction, on the $|j,m\rangle$ basis,
\begin{align}
    |v_{n,-}\rangle =\frac{1}{\sqrt{2}}\sum_{m=1}^j (c_{nm,-}|j,m\rangle -c_{nm,-}|j,-m\rangle),
\end{align}
therefore the Shannon entropy is
\begin{align}
    \bar{H}_n=-\sum_{m=1}^{j} |c_{nm,-}|^2\ln |c_{nm,-}|^2+\ln2.
\end{align}
It is then clear that $\bar{H}_n=H_n+\ln 2$, for $H_n\gg 1$, there is $H_n\simeq \bar{H}_n$. For Shannon ELM 
\begin{align}
    \bar{L}_n =\frac{e^{\bar{H}_n}}{2j+1}=\frac{e^{H_n}}{j+1/2}\stackrel{j\gg1}{\simeq}
\frac{e^{H_n}}{j} =L_n.
\end{align}
Similar results can be also proven for the even parity.
It demonstrates that for $j\gg1$, the Shannon ELM is the same whether within the parity subspace or not, while the Shannon entropy is almost the same if $H_n\gg1$. This implies that for the eigenvector statistics, particularly the mean Shannon ELM, if $j \gg 1$, it remains the same whether the Hilbert space is split or not. However, this is not the case for level statistics, where the Hilbert space must be split according to parity, because energy levels with different parity may not exhibit level repulsion.

Numerical result in Fig. \ref{fig:LocalLength} shows that $\langle L\rangle$ of $J_{x,-}$ in the basis $|j,m,-\rangle$ is close to the COE value. As we have proven, it is the same as the $\langle L\rangle$ of $J_x$ on $|j,m\rangle$ basis if $j\gg 1$ 
\begin{align}
    \langle L\rangle_{J_{x,-}}=\langle L\rangle_{J_x}= \frac{e^{\langle \bar{H}\rangle_{J_x}}}{2j+1},
\end{align}
where 
\begin{align}
    \label{eq:coe-jx}
&\langle \bar{H}\rangle_{J_x}=-\frac{1}{2j+1}\sum_{m,m'}|\langle j,m|j,m'\rangle_x|^2\ln|\langle j,m|j,m'\rangle_x|^2,\nonumber\\
     &=-\frac{1}{2j+1}\sum_{m,m'}|d_{m,m'}^j(\pi/2)|^2\ln|d_{m,m'}^j(\pi/2)|^2,
\end{align}
where $|j,m\rangle_x=\exp(-i\pi J_y/2)|j,m\rangle$. It means that $\langle L\rangle_{J_x}$ is equal to the mean Shannon ELM of the matrix $\exp(-iJ_y\pi/2)$ with each column as the eigenvector of $J_x$ in $|j,m\rangle$ basis. Numerical results show $\langle L\rangle_{J_x}$ is close to 0.48 at $j=2^{13}$, the Shannon ELM value of COE. The theoretical evaluation of $\langle L\rangle_{J_x}$ in Eq. \eqref{eq:coe-jx} and its closeness to the COE value is beyond the scope of this work.

The parity of the eigenvector is more clearly reflected in phase space through the spin coherent state $|\theta,\phi\rangle$, 
\begin{align}
    \langle \theta,\phi|v_{n,\pm}\rangle &= \langle \theta,\phi|R_x^2|v_{n,\pm}\rangle = \pm \langle \theta,\phi|R_x|v_{n,\pm}\rangle \nonumber \\ 
    &= \pm \langle \pi-\theta,-\phi|v_{n,\pm}\rangle,
\end{align}
where classically $R_x(x,y,z)^T=(x,-y,-z)^T$, and the parity is shown by eigenvectors in phase space just as sign difference. Regarding the splitting of parity, one could also define cat-state-like parity-symmetry-adapted coherent state \cite{calixto2012parity} $|\theta,\phi,\pm\rangle=(|\theta,\phi\rangle\pm|\pi-\theta,-\phi\rangle)/\sqrt{2}$. It is straightforward to see that this will yield the same results of Wehrl entropy and the localization measures.

\section{Mean ELM in integrable QKT}
\label{app-proof2}
\begin{figure}
    \includegraphics[width=0.9\linewidth]{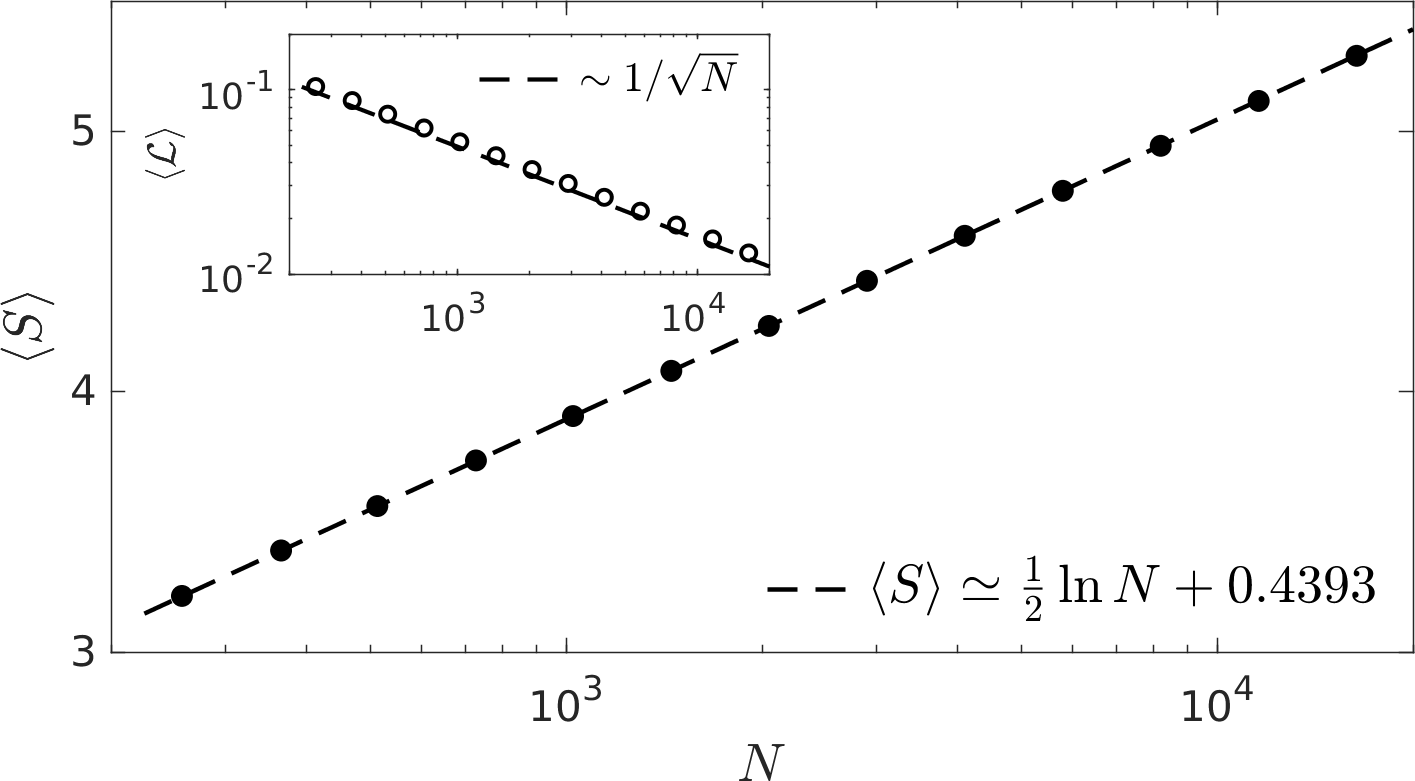}
    \caption{Mean Wehrl entropy $\langle S \rangle$, versus $N=2j+1$ for QKT with $\gamma=1$ and $\alpha=11\pi/19$. The inset is for mean Wehrl ELM $\langle \mathcal{L} \rangle$, calculated from $8\times 10^4$ coherent states held on $200\times 400$ grid points in phase space. The dashed lines indicate the best fittings.}
    \label{fig:integrable}
\end{figure}

\begin{figure}
    \includegraphics[width=0.85\linewidth]{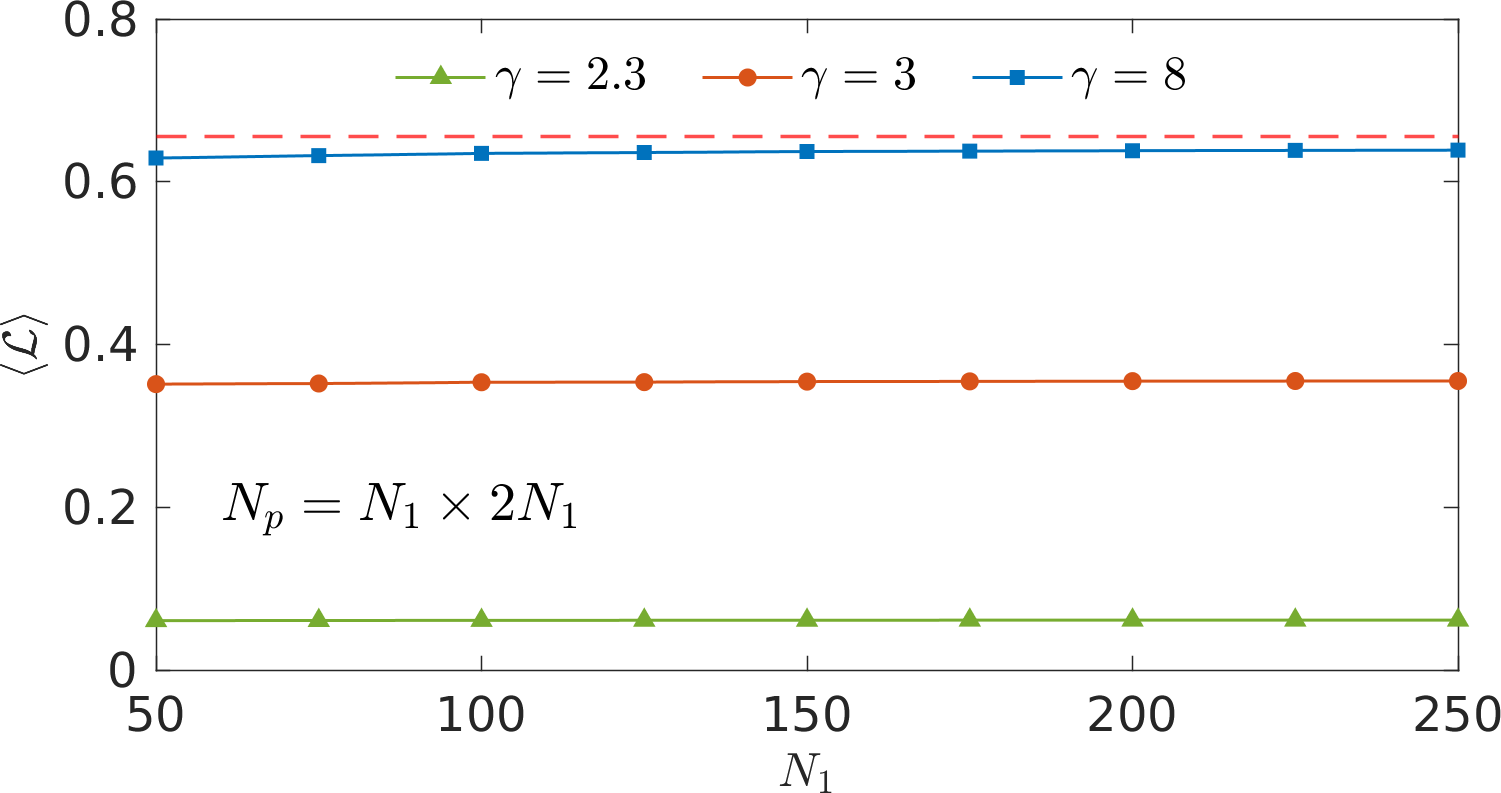}
    \caption{Mean Wehrl ELM versus the number of grid points in the $(\theta,\phi)$ phase space, with $N_p=N_1\times 2N_1$, for QKT with three different kicking strengths, with $j=2^{12}$, $\alpha=11\pi/19$. The red dashed line denotes the Wehrl ELM of CUE.}
    \label{fig:gridConverge}
\end{figure}

 In Fig. \ref{fig:LocalLength} we have shown that the mean Wehrl ELM $\langle \mathcal{L}\rangle$ of KT maintains a value equal to the mean Wehrl ELM of the unperturbed Hamiltonian $\langle \mathcal{L}\rangle_{J_x}$, for $\gamma\le 2$, which is almost integrable. Fig. \ref{fig:integrable} shows that at $\gamma=1$, the mean Wehrl entropy $\langle S\rangle \sim \frac{1}{2}\ln N$ and accordingly in the inset $\langle \mathcal{L}\rangle \sim N^{-1/2}$. When contrasting with the uniformly chaotic systems where $\langle S\rangle\sim \ln N$ and $\langle \mathcal{L} \rangle\sim 1$, the distinction between integrable and chaotic cases becomes apparent (see Fig. \ref{fig:qkt-rmt} for the behavior of $\langle S\rangle$ and $\langle \mathcal{L}\rangle$ at $\gamma = 8$). In integrable regimes, eigenstates tend to cover a thin band around a circle on the sphere, whereas in chaotic regimes, eigenstates exhibit delocalization across the entire sphere \cite{haake2013quantum}. We have examined $\langle \mathcal{L}\rangle$ with increasing number of grid points to $N_p=250\times 500$, for $j=2^{12}$. In Fig. \ref{fig:gridConverge} it shows that the mean Wehrl ELM is nearly unchanged versus the increasing number of grid points, for three different $\gamma$.

To derive the scaling of the mean Wehrl entropy of $J_x$,  it is first necessary to prove that $\langle S\rangle_{J_x}=\langle S\rangle_{J_z}$. While eigenvectors $|j,m\rangle_x$ of $J_x$ is a unitary transformation of eigenvectors of $J_z$, i.e. the Dicke basis $|j,m\rangle$, as
\begin{align}
    |j,m\rangle_x 
    =R(-\frac{\pi}{2},0)|j,m\rangle=\sum_{m'} d_{m'm}^j(\pi/2)|j,m'\rangle, 
\end{align}
where $R(\theta,\phi)$ is given in Eq. \eqref{scs-subeq1},  $d_{mm'}^j(\theta)$ is the Wigner $d$-matrix. 
The summation of the mean $q$-moment of Husimi functions of all the eigenvectors of $J_x$ can then be written as
\begin{align}
    \label{eq:app2-1}
   &\frac{1}{N}\sum_m W_{|m\rangle_x}^{(q)}
=\frac{1}{4\pi}\sum_m\int d\Omega |\langle \theta,\phi|j,m\rangle_x|^{2q} \nonumber \\
 &=\frac{1}{4\pi}\sum_m\int d\Omega\left|\langle j,j|R^\dagger(\theta,\phi)R(-\frac{\pi}{2},0)|j,m\rangle\right|^{2q}, 
\end{align}
 where $d\Omega=\sin\theta d\theta d\phi$, and $N=2j+1$.   Following Eq. \eqref{scs-subeq2}, we can further expand Eq. \eqref{eq:app2-1} with the properties of the Wigner $d$-matrix: $d_{mm'}^j(\theta)=d_{m'm}^j(-\theta)=(-1)^{m'-m}d_{m'm}^j(\theta)$, for integer $q$. It yields
 \begin{align}
  &\frac{1}{4\pi}\sum_m\int d\Omega \big|(1+|\xi|^2)^{-j}\sum_{m'}\binom{2j}{j+m'}^{\frac{1}{2}}\xi^{j-m'}d_{m'm}^j(\frac{\pi}{2})\big|^{2q}\nonumber \\
  &= \frac{1}{4\pi}\sum_m\int d\Omega \big|(1+|\xi|^2)^{-j}\sum_{m'}\binom{2j}{j+m'}^{\frac{1}{2}}\xi^{j-m'}\big|^{2q}\nonumber\\
  &=\frac{1}{4\pi}\sum_m\int d\Omega |\langle \theta,\phi|j,m\rangle|^{2q}=\frac{1}{N}\sum_m W_{|m\rangle}^{(q)}.
 \end{align}

 Considering the equivalence of mean $q$-moments, we establish that $\langle S\rangle_{J_x}=\langle S\rangle_{J_z}$, as $S^{(q)}_{|\psi\rangle} =\frac{1}{1-q} \ln W^{(q)}_{|\psi\rangle}$. While for Dicke states, it was proven in the limit $q\to 1$ \cite{lee1988wehrl}
 \begin{align}
   & S_{|m\rangle}=\frac{2j}{2j+1}-\ln\binom{2j}{j-m}+2j\Psi(2j+1)\nonumber\\
   &-(j+m)\Psi(j+m+1)-(j-m)\Psi(j-m+1),
 \end{align}
where $\Psi(x)$ denotes the digamma function, and obviously $S_{|m\rangle}=S_{|-m\rangle}$. Therefore,
\begin{align}
    \langle &S\rangle_{J_z}= \frac{1}{N} \sum_{m=-j}^j S_{|m\rangle} = \frac{2j}{2j+1}-\ln (2j)! +2jB_{2j}\nonumber \\
    &+ \frac{2}{2j+1}\sum_m\left[\ln (j+m)!-(j+m)B_{j+m}\right],
\end{align}
where $B_n=\sum_{k=1}^n \frac{1}{k}\approx \ln n +\gamma_e$. Applying Stirling's approximation to factorials $\ln n!\approx n\ln n +\frac{1}{2}n-n$, for $j\to \infty$, one then gets $ \langle S\rangle_{J_z}\sim \frac{1}{2}\ln N$. Consequently, the mean ELM $\langle \mathcal{L} \rangle$  scales as $N^{-1/2}$,  the same scaling as the width of effective Planck cell $j^{-1/2}\propto N^{-1/2}$.

\section{Gallery of states}
\label{app-gallery}
\begin{figure*}
    \includegraphics[width=0.8\linewidth]{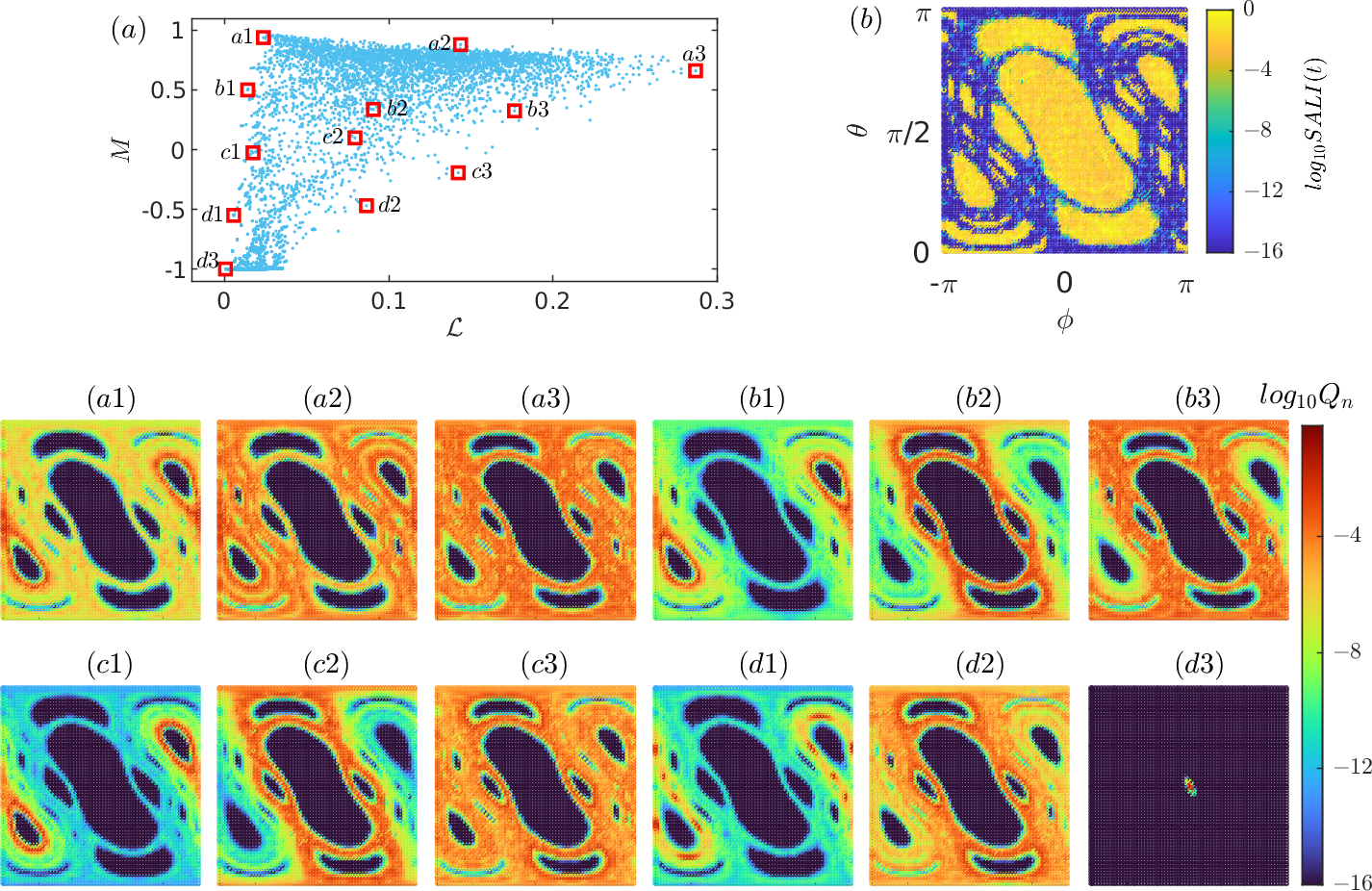}
    \caption{Analogous data as in Fig. \ref{fig:husimi}, but for $\gamma=2.3$. The darkest blues in the plots of Husimi functions from (a1)-(d3) show the area where $Q_n< 10^{-16}$. The most localized state (d3) precisely located at the (trivial) fixed point $(\theta,\phi)=(\pi/2,0)$, while the state (a1) of the maximum overlap index, is strongly localized around and avoids the outer tori of the regular region. }
    \label{fig:husimi1}
\end{figure*}

\begin{figure*}
    \includegraphics[width=0.8\linewidth]{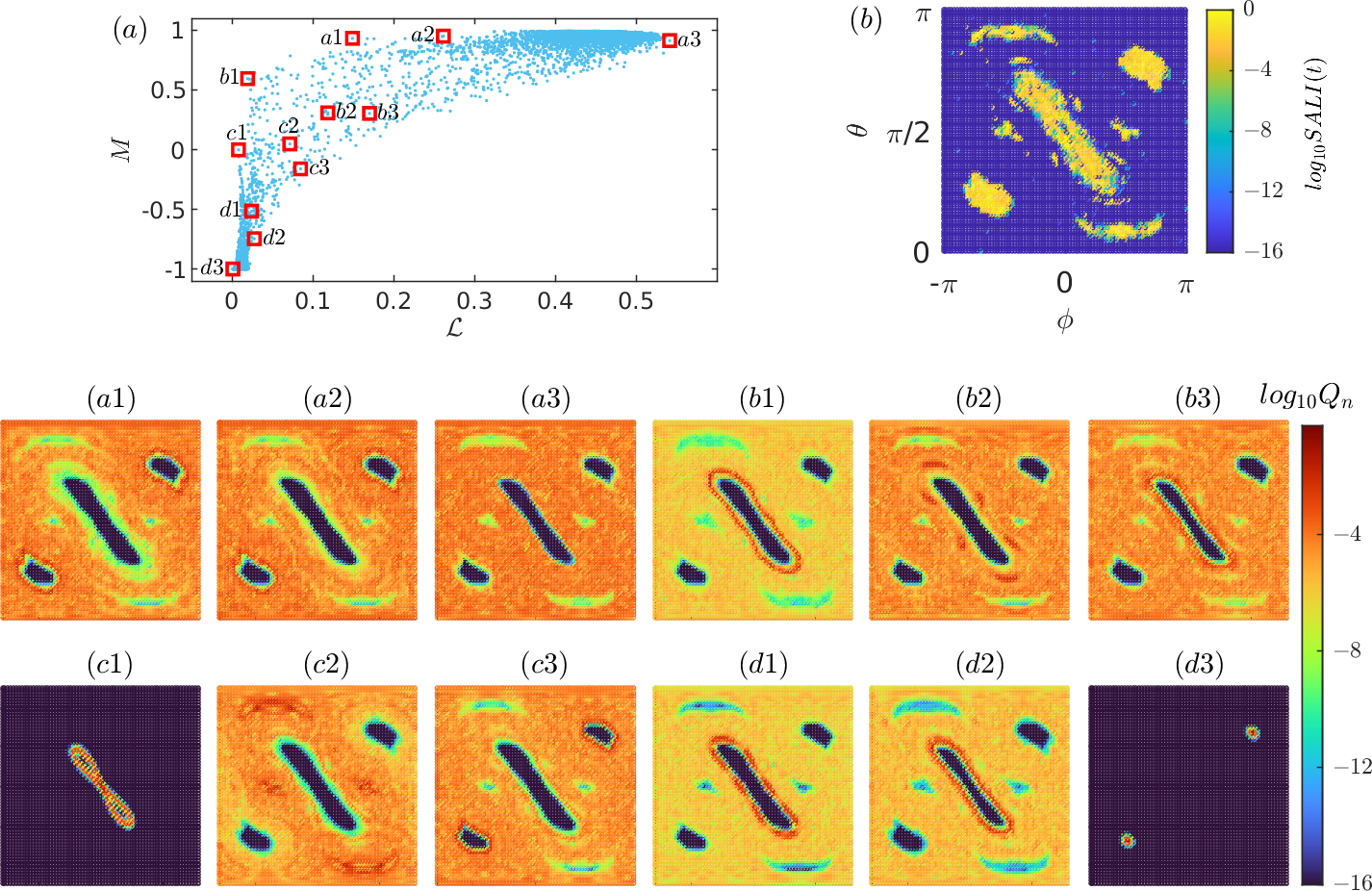}
    \caption{Analogous data as in Fig. \ref{fig:husimi} and  \ref{fig:husimi1}, but now for $\gamma=3$. Compared with $\gamma=2.3$, in this case, the fixed point at $(\theta,\phi)=(\pi/2,0)$ becomes unstable, which is clearly shown in the SALI plot. Interestingly, the most localized state (d3) occupies identical fixed points, akin to when $\gamma=2.6$.}
    \label{fig:husimi2}
\end{figure*}

Here, we provide additional examples of the joint distribution $(\mathcal{L}, M)$ for mixed-type QKT, along with representative Husimi functions of eigenstates from different regions of the parameter space, 
at two kicking strengths $\gamma=2.3$ (Fig. \ref{fig:husimi1}) and $3$ (Fig. \ref{fig:husimi2}), accompanied by corresponding SALI plots.

\bibliography{addendum}
\end{document}